\documentclass[12pt, a4paper]{iopart}

\usepackage{iopams}  
\usepackage[dvips]{graphics}
\usepackage{graphicx}
\usepackage{braket}
\usepackage{bm}
\usepackage{dsfont}

\usepackage{color}

\begin{document}
\title[Direct probing of the Wigner function ...]{Direct probing of the Wigner function by time-multiplexed detection of photon statistics}

\author{K Laiho, M Avenhaus, K N Cassemiro, and Ch Silberhorn}

\address{Max Planck Institute for the Science of Light, G\"unther-Scharowsky-Stra\ss e 1/Building 24, 91058 Erlangen, Germany}
\ead{Kaisa.Laiho@mpl.mpg.de}

\begin{abstract}
We investigate the capabilities of loss-tolerant quantum state characterization using a photon-number resolving, time-multiplexed detector (TMD). We employ the idea of probing the Wigner function point--by--point in phase space via photon parity measurements and displacement operations, replacing the conventional homodyne tomography. Our emphasis lies on reconstructing the Wigner function of non-Gaussian Fock states with highly negative values in a scheme that is based on a realistic experimental setup. In order to establish the concept of loss-tolerance for state characterization we show how losses can be decoupled from the impact of other experimental imperfections, i.e. the non-unity transmittance of the displacement beamsplitter and  non-ideal mode overlap. We relate the experimentally accessible  parameters to effective ones that are needed for an optimised state reconstruction. The feasibility of our approach is tested by Monte Carlo simulations, which provide bounds resulting from statistical errors that are due to limited data sets. Our results clearly show that high losses can be accepted for a defined parameter range, and moreover, that---in contrast to homodyne detection---mode mismatch results in a distinct signature, which can be evaluated by analysing the photon number oscillations of the displaced Fock states.
\end{abstract}

\pacs{03.65.Wj, 42.50.-p, 42.50.Ar}
\vspace{2pc}
\section{\label{intro}Introduction}
The preparation and accurate characterization of non-Gaussian states is of paramount importance for quantum processing applications.  For example the implementation of scalable continuous variable quantum information protocols require an entanglement distillation procedure, which has been proven to be impossible using only a set of Gaussian states and Gaussian operations~\cite{J.Eisert2002, J.Fiurasek2002, G.Giedke2002}.
The most fundamental {\it non-Gaussian} states are Fock states.  Recent experiments have demonstrated for the first time the generation of Fock states  and photon added/substracted states~\cite{A.I.Lvovsky2001, A.Zavatta2004, J.S.Neergaard-Nielsen2006, A.Ourjoumtsev2006, K.Wakui2006}. In all of these experiments conventional homodyne detection~\cite{H.P.Yuen1983, B.N.Schumaker1984} followed by tomographic reconstruction was employed to prove the presence of the non-Gaussian characteristics, as well as the presence of negative values, in the Wigner functions of the quantum states~\cite{Vogel1989}. These negative values are considered to be a signature of quantum behaviour.
 In a homodyne measurement, the detected signal consists of a beating between the quantum signal and a strong local oscillator field, which in turn has the effect that mode mismatch cannot be distinguished from losses. Another drawback in this method is the fact that homodyne tomography is always an indirect state characterization: the Wigner function can be reconstructed only after intricate computational back-projection, and the photon statistics have to be recovered by more involved means~\cite{D.T.Smithey1993,U.Leonhardt1995,Lvovsky2005}. 

A more direct characterization of the Wigner function is possible if a photon number resolving detector is available. The Wigner function at a given position $\alpha$ of phase space ($\alpha \in \mathbb{C}$) is equal to the average value of the photon number parity  operator  on a quantum state displaced by  $-\alpha$~\cite{K.E.Cahill1969}. The expectation value of the parity operator is directly inferred from a complete measurement of the photon number distribution~\cite{S.Wallentowitz1996, K.Banaszek1996}. As opposed to tomography, in this method the Wigner function is probed point--by--point in phase space, resulting in  a high resolution measurement. Until now, the requirement of measurements with photon number resolution has severely limited the usage of this method, although different ideas to overcome the problem have arisen for specific systems~\cite{K.Banaszek1999, Lutterbach1997, P.Bertet2002}.

Recently, new strategies for measuring light with photon number resolution have been developed, which allow a more intensive exploration and characterization of quantum states~\cite{Silberhorn2007}. One of them---applicable for pulsed light---is a time multiplexed detector (TMD). 
 With the TMD, the photon number distribution of a quantum state can be revealed via an inversion procedure applied to the measured photon statistics. Experimental applications, demonstrating a reliable loss calibration, and the TMD's suitability for detecting multimode statistics have been accomplished~\cite{D.Achilles2005,M.Avenhaus2008}. 
 
 In this paper, we apply for the first time to our knowledge the probing method in connection with a TMD to reconstruct the Wigner functions of single-photon and two-photon Fock states, showing the capability of loss-tolerant {\it state} characterization.
 This is achieved by decoupling the effects of experimental imperfections from the ideal mathematical operations. We find an equivalent model for the experimental setup, in which losses are independent from the displacement operation. In addition,  mode mismatch is modeled by a convolution of a Poissonian distribution with the statistics of an ideally displaced Fock state.
Via Monte Carlo simulations we evaluate the specific experimental conditions necessary to characterize reliably non-Gaussian Fock states. Using a reasonable number of data points ($10^6$) we show the extent of degradation caused by losses and mode mismatch. 
A simple inspection of the characteristic photon number oscillation of displaced Fock states enables us to quantify the degree of overlap.

This paper is organised as follows. 
In Sec.~\ref{BSsec} we briefly introduce the method for probing the Wigner function using displacement operations and photon counting. 
Experimental issues related to the implementation of the ideal displacement operation are discussed in Sec.~\ref{displacementSec}.
In Sec.~\ref{TMDsec} we recall important properties of a TMD and explain how to handle the effect of losses via an inversion algorithm.
In Sec.~\ref{modeoverlapSec} we introduce a model to study the effect of imperfect mode overlap, which is used in Sec.~\ref{diagnosis} in order  to relate the degree of overlap to usual experimental observables. 
In the following section (Sec.~\ref{highlosses}),  we exploit this theoretical framework and employ Monte Carlo simulations to analyse the reliability of the probing technique under the usual circumstances of high losses and imperfect mode overlap. We show how  the Wigner function characterictics can be efficiently recovered from the response of the TMD.  
Finally, in  Sec.~\ref{conclusions} we conclude and highlight the main findings of our investigations.

\section{\label{BSsec} Wigner function reconstruction by direct probing method}
The probing method is based on a simple expression that relates the Wigner function at the origin of the phase space 
with the mean value of the photon number parity  operator $\hat{\Pi}$~\cite{S.Wallentowitz1996, K.Banaszek1996},
\begin{equation} 
	W(0) = \frac{2}{\pi}\;\mathrm{Tr}(\,\hat{ \rho}\, \hat{\Pi}\,).
	\label{W_origo}
\end{equation}
As the Fock states  $\ket{n}$ are eigenstates of the photon number parity operator, with eigenvalues $(-1)^n$, a natural representation for the density matrix is $\hat{\rho}=\sum {\rho}_{mn}\ket{m}\bra{n}$. With this representation Eq.~(\ref{W_origo}) simplifies to 
\begin{equation}
	W(0) = \frac{2}{\pi}\sum_{n}(-1)^{n}\rho_{nn}\,.
	\label{W_0}
\end{equation}
Since $\rho_{nn}$ is equal to the probability of counting $n$ photons, the Wigner function can be directly measured from the photon  statistics.

Due to the fact that the projection onto the photon number basis always erases the information about the coherence of the state, the parity measurement returns the value of the Wigner function only  at one single  point of phase space. Nevertheless, the whole Wigner function can be reconstructed if appropriate displacements are applied to the state. To access its value at a given position $\alpha$ in phase space we need to displace the state by the quantity $-\alpha$. Mathematically, this is described by the action of the displacement operator 
$\hat{D}(-\alpha)=\hat{D}^{\dagger}(\alpha)$. The evolution of the density operator is given by $\hat{\rho}^{\prime}(-\alpha)=\hat{D}(\alpha)^{\dagger}\hat{\rho} \hat{D}(\alpha)$, which results in 
\begin{equation}  
	W(\alpha) = \frac{2}{\pi}\;\mathrm{Tr} (\,\hat{D}^{\dagger}(\alpha)\hat{\rho} \hat{D}(\alpha)\, \hat{\Pi}\, )=  \frac{2}{\pi}\sum_{n}(-1)^{n}\rho^{\prime}_{nn}(-\alpha)\,.
	\label{W_pn}
\end{equation}
Similarly  to the expression in Eq.~(\ref{W_0}), the value of the Wigner function in  Eq.~(\ref{W_pn}) is determined directly by the photon statistics of the displaced state.

In general terms, the reconstruction of a quasi-probability distribution function of a degraded quantum state via the direct probing technique has been analysed in the past~\cite{S.Wallentowitz1996,K.Banaszek1997, K.Banaszek2002}. 
Here, we extend the previous work by showing, how to deal with the experimental imperfections in practical terms in order to achieve a loss-tolerant state characterization.
To increase the amount of information, which can be extracted from  our simulations  we establish a strong connection between the theory and the actual experimental parameters that constrain the reconstruction of the Wigner function. 
In the course of this work we evaluate the impact of imperfections in this characterization method, i.e.~losses and mode mismatch between fields. In addition, an important degrading effect comes from the implementation of the displacement, which nevertheless can be handled with a loss inversion strategy similar to the inefficient detection.


\section{\label{displacementSec}Experimental implementation of the displacement operation}
\begin{figure}[!ht]
 	\begin{center}
		\includegraphics[width= 0.75\textwidth]{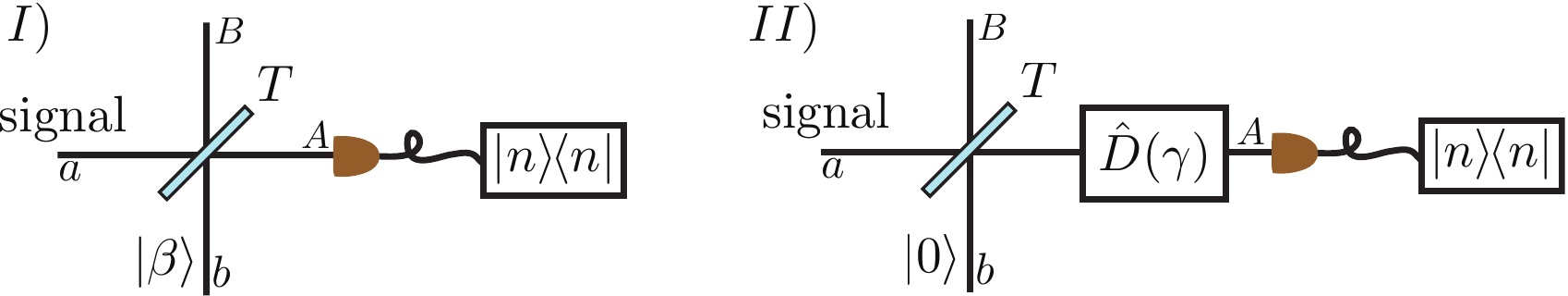}
		\caption{I) The displacement operation can be experimentally realized by combining a quantum signal with a coherent reference field in a highly asymmetrical beam splitter.  II) Sketch of a mathematical model that predicts exactly the same photon number distribution as the model shown on the left side of the picture. The equivalence is achieved when $\gamma = \sqrt{1-T}\,\beta$. }
		\label{displacement}
	\end{center}
\end{figure}

In an experiment the displacement operation  can be well approximated by the effect of a highly asymmetric beam splitter with transmission $T\approx 1$, which is used to superimpose the quantum signal of interest with a  coherent reference field $\ket{\beta}$~\cite{G.M.D'Ariano1995, M.G.A.Paris1996}. 
The beam splitter is a linear optical element that connects two input modes $a$ and $b$, delivering output modes $A$ and $B$, as shown in Fig.~\ref{displacement}(I). 
In the Heisenberg picture, the output annihilation operators are related to input operators by the transformations
\begin{eqnarray}
\hat{A}  = \sqrt{T}\hat{a}+\sqrt{1-T}\hat{b}\,, 
\label{beamsplitter}\\
\hat{B}= \sqrt{T}\hat{b}-\sqrt{1-T}\hat{a}\,,
\label{beamsplitter2}
\end{eqnarray}
with $T$ and $R=1-T$ being equal to the transmission and reflection intensity coefficients. 
From the experimental point of view, the beam splitter will always introduce some amount of loss, since $T \neq 1$. To decouple the displacement from the loss we analyse the configuration shown in Fig.~\ref{displacement}(II). 

It is straightforward to derive the conditions under which the networks (I) and (II) are equivalent, if we follow an approach introduced by  Wallentowitz and Vogel~\cite{S.Wallentowitz1996}. Firstly, using the beam splitter transformation in Eq.~(\ref{beamsplitter}), the output annihilation  operators of each model are related to the input modes by,
\begin{eqnarray}
\hat{A}^{(I)} &=& \sqrt{T}\hat{a}+ \sqrt{1-T}\hat{b}\,,
\label{network1II}\\
\hat{A}^{(II)} &=& \sqrt{T}\hat{a}+ \sqrt{1-T}\hat{b} + \gamma\,, 
\label{network1I}
\end{eqnarray}
in which we employed the relationship $D^{\dagger}(\gamma)\hat{A} D(\gamma)=\hat{A}+\gamma$.

Second, we note that in a photon-counting experiment the probability of recording $n$ photons  is given by~\cite{L.Mandel1995} 
\begin{eqnarray}
P_{n} = \braket{: e^{-\hat{N}}\;\frac{\hat{N}^n }{n!}:} 
\label{P_N}
\end{eqnarray}
with the symbol $::$ defining a normal ordering of the operators. The average in Eq.~(\ref{P_N}) is taken over the initial quantum state $\ket{\Psi}$ and $\hat{N}=\hat{A}^{\dagger}\hat{A}$ is the number operator of the detection mode.

As mode $b$ is prepared in a coherent state in both models, and $P_{n}$ is evaluated by the average of normally ordered operators, the operator $\hat{b}$ can be replaced by a complex number $\beta$, with $\hat{b}\,\ket{\beta}=\beta\,\ket{\beta}$. Thus, the annihilation  operators from models (I) and (II) introduced by the number operator in Eq.~(\ref{P_N}) can be substituted by
\begin{eqnarray}
\hat{A}^{(I)} &\rightarrow& \sqrt{T}\hat{a}+ \sqrt{1-T}\beta=\sqrt{T}\left(\hat{a}+ \sqrt{\frac{1-T}{T}}\,\beta\right)\,,\\
 \hat{A}^{(II)} &\rightarrow& \sqrt{T}\hat{a}+ \gamma=\sqrt{T}\left(\hat{a}+ \frac{1}{\sqrt{T}}\,\gamma \right)\,.
\end{eqnarray}
We are only interested in the photon number probability $P_{n}$ and therefore define effective photon number operators $\hat{N}^{(I)}_{\mathrm{eff}}$ and $\hat{N}^{(II)}_{\mathrm{eff}}$, which are independent of losses but represent effective displacements. The number operators for networks (I) and (II) can then be substituted by
\begin{eqnarray}
\hat{N}^{(I)} &\rightarrow& T\,\hat{N}^{(I)}_{\mathrm{eff}} = T\hat{D}^{\dagger}\left(\sqrt{(1-T)/T}\;\beta \right) \; \hat{a}^{\dagger}\hat{a} \; \hat{D} \left(\sqrt{(1-T)/T}\;\beta \right)  \,, \\
\hat{N}^{(II)} &\rightarrow&  T\, \hat{N}^{(II)}_{\mathrm{eff}} =  T\hat{D}^{\dagger}\left(\sqrt{1/T}\gamma \right)\;\hat{a}^{\dagger}\hat{a}\; \hat{D} \left(\sqrt{1/T}\gamma \right) \,.
\label{NII}
\end{eqnarray}
Examining the last two equations we reason that both schemes manifest equal photon number probabilities if $\gamma=\sqrt{1-T}\beta$. We would like to emphasise that the loss element in Eq.~(\ref{NII}) modelled by the beam splitter with transmission coefficient $T$, increases the net amount of displacement, i.e. the {\it total} amount of displacement is $\alpha=\gamma / \sqrt{T}$. In addition to the displacement, the resulting probability is given by an attenuated signal with an the overall quantum efficiency of $\eta= T$. As a conclusion we find that even when the beam splitter transmission is far from the idealized case, the device still works as a displacing element and introduces simply some amount of extra loss that can be inverted as will be described in Sec.~\ref{TMDsec}.

To simulate the possible statistics resulting from an actual  measurement, we can use network (II) to calculate the photon number distributions of displaced Fock states $\hat{D}(\gamma)\ket{m}$. In the absence of loss, these photon number distributions are well known and given by $\rho_{nn}(\gamma)= |\braket{n|\hat{D}(\gamma)|m}|^{2}$~\cite{F.A.M.Olivera1990}. The effect of losses can be analysed in a very straightforward fashion. 
For Fock states a beam splitter with transmission coefficient $T$ yields a statistical mixture, which is governed by  binomial distribution
\begin{eqnarray}
	P^{\hat{D}\ket{m}}_{n} = \sum_{j=0}^{m} \left( {\begin{array}{*{20}c} m\\ j\\ \end{array}} \right) T^{j}(1-T)^{m-j} \left | \braket{n | \hat{D}(\gamma) | j} \right |^{2}\,.
	\label{idealD}
\end{eqnarray}
For the Fock states $\ket{1}$ and $\ket{2}$  this results explicitly in 
\begin{eqnarray}
	\hspace{-60pt}P^{\hat{D}\ket{1}}_{n} = 
	e^{-\left |\gamma\right |^{2}}\frac{|\gamma|^{2n}}{n!} \left [1-T+ \frac{T\left (n-|\gamma|^{2}\right)^{2}}{|\gamma|^{2}}\right ], \;\;\;\;\mathrm{} \,  \nonumber 
	\label{P_displaced1}
\end{eqnarray}
\begin{eqnarray}
\hspace{-60pt}	P^{\hat{D}\ket{2}}_{n} = 
e^{-\left |\gamma\right |^{2}}\frac{|\gamma|^{2n}}{n!} \left [(1-T)^{2}+ 2(1-T) T\frac{\left (n-|\gamma|^{2}\right)^{2}}{|\gamma|^{2}} + T^{2} \frac{  \left ( (n-|\gamma|^{2})^{2}-n \right)^{2}}{2|\gamma|^{4}}\right]\,\,.
	\label{P_displaced2}
\end{eqnarray}

\section{\label{TMDsec}Loss tolerant characterization using TMD}

\begin{figure}[!ht]
	\begin{center}
		\includegraphics[width= 0.7\textwidth]{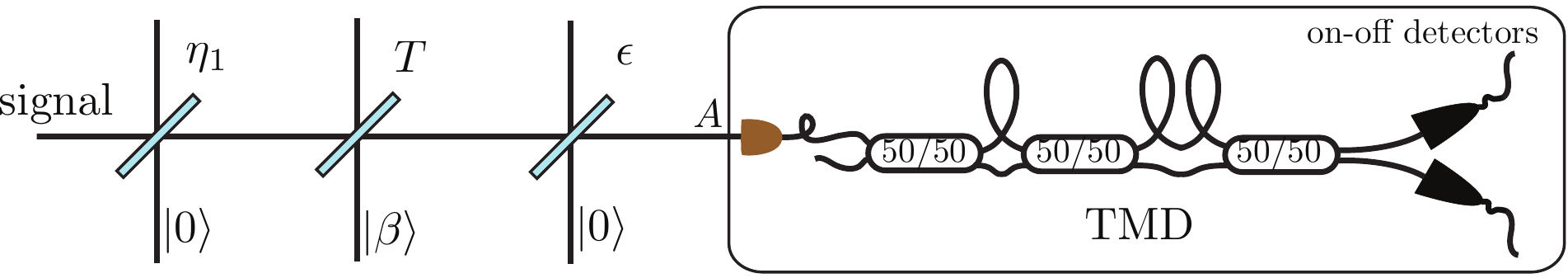}
		\caption{Representation of a theoretical model in which an input quantum signal undergoes a loss, followed by a non-ideal displacement and is finally measured with inefficient detection. The displacement is realised by a high-transmission beam splitter in which the signal is superimposed with a weak reference field  and the photon number resolution is achieved with a TMD. 
		}
		\label{scheme}
	\end{center}
\end{figure}

According to the previous Sec.~\ref{displacementSec} the measurable probability $P_{n}$ differs from the ideal $\rho^{\prime}_{nn}$ that is required for the reconstruction of the Wigner function in Eq.~(\ref{W_pn}). 
In this section we study how to relate these quantities and generalise the model to allow for the detection losses. In this case the quantum signal is degraded both before and after the realisation of the displacement. 
The network shown in  Fig.~\ref{scheme} represents the sequence of operations undergone by the signal, where losses are modelled as beam splitters combining the signal with vacuum modes. The last beam splitter, with transmission $\epsilon$, represents the limited quantum efficiency of the detector itself. 
As before, to evaluate the photon number probabilities, we introduce the effective photon number operator, which can be expressed as
\begin{eqnarray}
\hat{N}&\rightarrow&  \epsilon T \eta_1 \hat{N}_{eff}= \epsilon T \eta_1 \hat{D}^{\dagger}\left(\sqrt{1/(\eta_1\,T)}\gamma \right)\;\hat{a}^{\dagger}\hat{a}\; \hat{D} \left(\sqrt{1/(\eta_1\,T)}\gamma \right) \,,
\label{Ntmd}
\end{eqnarray}
with $\eta=\epsilon T \eta_1$ being the overall quantum efficiency. The total amount of displacement is equal to $\alpha=\gamma / \sqrt{\eta_1 T}$. Note that in contrast to losses  experienced by the signal before the displacement, the detection efficiency $\epsilon$ has no impact on $\alpha$.

In analogy to Eq.~(\ref{P_N})  with the detection loss $\eta$, we note that the exponential factor $-\eta \hat{A}^{\dagger}\hat{A}$ can be rewritten as $[1+(1-\eta)]\hat{A}^{\dagger}\hat{A}$. 
Employing the series expansion of exponential function it is easy to show that
\begin{eqnarray}
P^{\hat{D}\ket{\Psi}}_{n} &=& \braket{:\sum_{j=n}^{\infty}  \left( {\begin{array}{*{20}c} j \\ n \\ \end{array}} \right) \eta^n (1-\eta)^{j-n} e^{-(\hat{N}_{eff})}\;\frac{(\hat{N}_{eff})^j }{j!}:}  \nonumber \\
&=&  \sum_{j=n}^{\infty} \left( {\begin{array}{*{20}c} j \\ n \\ \end{array}} \right) \eta^n (1-\eta)^{j-n} 
\braket{:e^{-(\hat{N}_{eff})}\;\frac{(\hat{N}_{eff})^j }{j!}:} \nonumber \\
&=&  \sum_{j=n}^{\infty} L_{n,j} |\braket{j|\hat{D}|\Psi}|^{2}\, = \sum_{j=n}^{\infty} L_{n,j} \rho^{\prime}_{jj},
\label{P_Nloss}
\end{eqnarray}
where we define $L_{n,j}$ as an element of an upper diagonal matrix, referred as the loss matrix.
Recording $n$ events with a non-ideal detection, ensures that the signal contained at least $j\geq n$ photons. We note that Eq.~(\ref{P_Nloss}) differs from Eq.~(\ref{idealD}) in the sense that the right hand term $|\braket{j|\hat{D}(\alpha)|\Psi}|^{2}$ directly  corresponds to the non-degraded statistics of the displaced initial state $\hat{D}(\alpha)\ket{\Psi}$.  If Eq.~(\ref{P_Nloss}) is expressed in matrix form $\vec{P}^{\hat{D}\ket{\Psi}}=\mathbf{L}(\eta)\,\vec{\rho}^{\;\hat{D}\ket{\Psi}}$, we can easily verify that the loss inverted statistics correspond to the ideal displaced statistics. Thus, we can deduce that  the effect of non-ideal detection can in principle be handled by the loss inversion method. Nevertheless,  photodetectors with photon number resolution are still essential. 

Photon number resolution can be achieved  by distributing several photons into many different temporal modes, which are then analysed with binary detectors by time multiplexing. In practice, a TMD consists of a fibre-integrated 50/50 beam splitter network with time delays implemented as fibre loops of different lengths (Fig.~\ref{scheme}). A pulse launched into the detector is divided into a train of pulses in two separate fibre outputs~\cite{M.Fitch2003, D.Achilles2003}. The existence of photons in each mode (or time-bin) can be detected by  avalanche photo diodes (APDs) and the so-called click-statistics are collected for ensemble measurements.

The measured click statistics $\vec{p}=(p_0,\,p_1,\,...\,p_n)$ are related to the input photon number distribution 
through the relation~\cite{D.Achilles2004}
\begin{equation}
\vec{p}=\mathbf{C}\vec{P}^{\hat{D}\ket{\Psi}}=\mathbf{CL}(\eta)\,\vec{\rho}^{\;\hat{D}\ket{\Psi}}\,.
\label{CL_inversion}
\end{equation}
The convolution matrix $\mathbf{C}$ takes into account a stochastic distribution of $n$ photons into several bins (from which follows that the number of clicks $N$ is lower or equal to $n$) and the matrix $\mathbf{L}(\eta)$ describes the binomial process of loss,
as introduced in Eq.~(\ref{P_Nloss}). Therefore, by inverting equation~(\ref{CL_inversion}) (or using a maximum likelihood technique), the photon number distribution of the quantum signal can be reconstructed from the click statistics if the overall loss $\eta$ and the convolution matrix are known.  Beside the numerical techniques an analytical solution   can be used for inverting the losses, as presented in \ref{App_A}.

Experimentally, the calibration of the overall efficiency $\eta=\eta_1 T \epsilon$ can be reliably accomplished by  taking advantage of the twin photon generation properties of a parametric down conversion process~\cite{D.Achilles2005, M.Avenhaus2008}. 
In addition to the overall loss,  the total amount of displacement $\alpha = \gamma/\sqrt{\eta_1 T} =  \sqrt{1-T} \beta /\sqrt{\eta_1 T} $  needs to be properly calibrated. 
The transmission $T$ of the beam splitter can be characterized by monitoring both outputs of the beam splitter and
measuring the fraction of a strong input field $\beta$ that is transmitted and detected by a standard photo diode.
An additional measurement extracts the value of $\epsilon$ by, e.g. measuring the  Poissonian statistics of the coherent reference field $\ket{\beta'} = \ket {\sqrt{(1- T) \epsilon }\beta}$ that is incident on the TMD.
Finally, combining the results of the unbalanced homodyning of vacuum together with the knowledge of the overall losses and $T$, enables us to determine the amount $\eta_1$. Thus, all losses can be experimentally characterized enabling the loss-tolerant probing of the Wigner function.

\section{\label{modeoverlapSec}Imperfect mode overlap between quantum signal and reference beam}

Mode overlap is a crucial issue for both unbalanced and balanced characterization techniques. In balanced homodyne measurements the detected photocurrent $\hat{I}$ is proportional to the {\it product} of the quantum signal and the local oscillator, $\hat{I}\propto \hat{a}\,\beta^* + \hat{a}^{\dagger} \beta$, \cite{Lvovsky2005}.  As only the overlapping part of the signal state is measured, imperfect mode overlap has intrinsically the same signature as a loss. Contrariwise, in unbalanced measurements such as the direct probing method, the detection is proportional to the photon number operator, represented by {\it sum} of the signal and reference field, i.e.~$\hat{n} \propto (\hat{a}^{\dagger}+\beta^*)(\hat{a}+ \beta)$. Each photon impinging on the APD causes a detectable photocount, and   the signature of the mode mismatch becomes different from loss \cite{footnotea}.



\begin{figure}[!ht]
	\begin{center}
		\includegraphics[width= 0.65\textwidth]{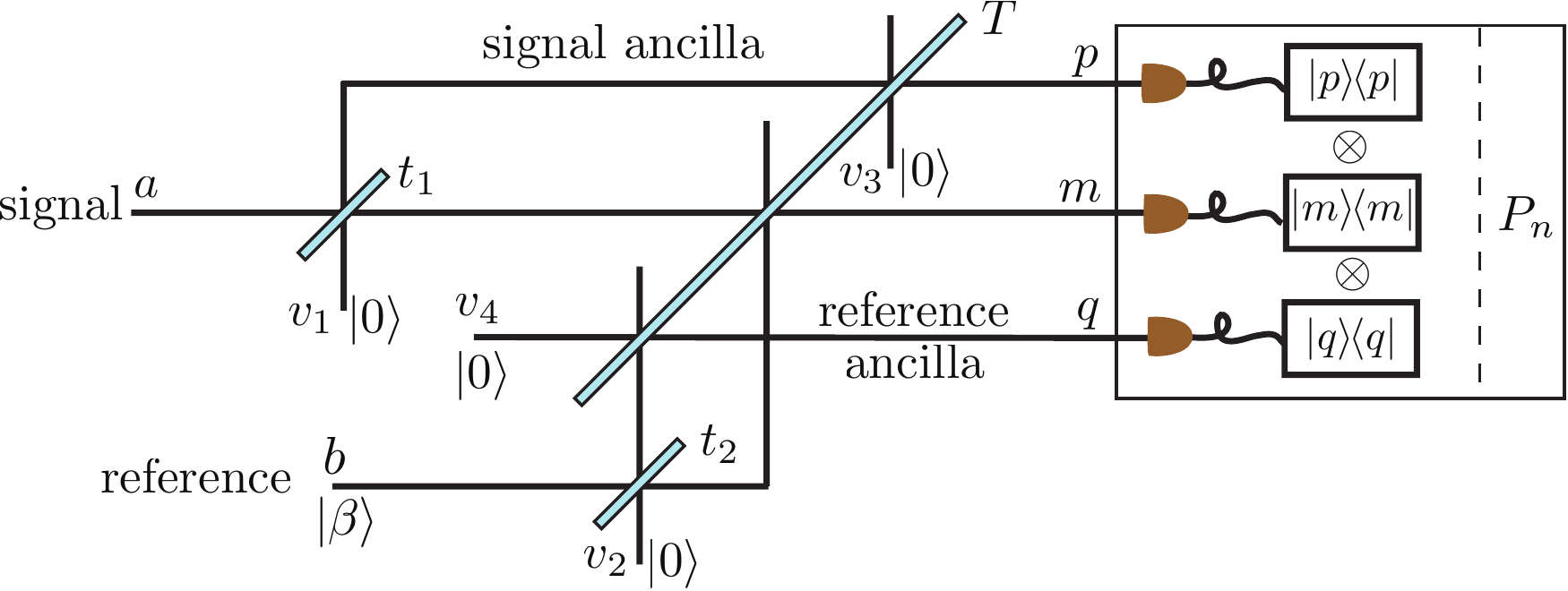}
		\caption{Network  to model the imperfect mode overlap. Signal and coherent reference field are divided into four modes. 
		The displacement occurs at the beam splitter with transmittance $T$. Each detector measures a different contribution, which are convolved to observe the joint distribution $P_n$.
		}
		\label{3BS_overlap}
	\end{center}
\end{figure}

\begin{figure}[!ht]
	\begin{center}
		\includegraphics[width= 0.35\textwidth]{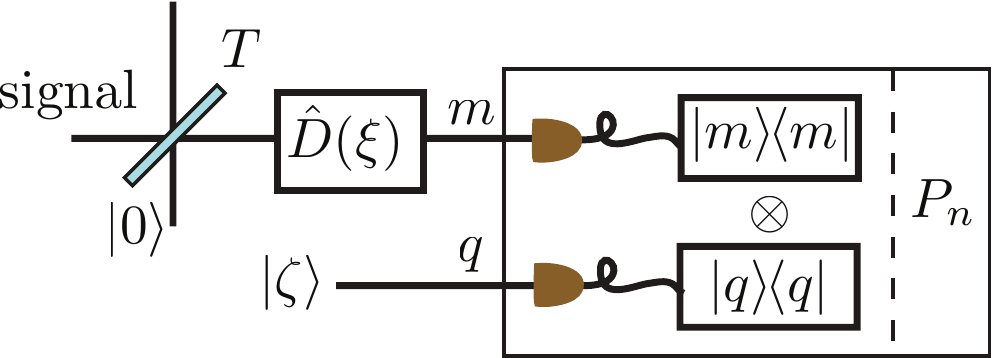}
		\caption{Simplified model of mode overlap with  two independent modes. For details see text.		}
		\label{simple_overlap}
	\end{center}
\end{figure}

We model the imperfect mode overlap by the beam splitter network as shown in Fig.~\ref{3BS_overlap}.
In the following we elaborate the conditions under which this network is equivalent to the one depicted in Fig.~\ref{simple_overlap}. As before we relate the actual experimental parameters with the values of the equivalent network. 
In the network of Fig.~\ref{3BS_overlap}, the signal and the reference beam are divided into four  modes, two of them interfering. The other modes do not overlap and are just split according to the ratio $T$.
Hereby we assume that $t_{1}$ and $t_{2}$ correspond to the fractions of the signal and reference fields, which are perfectly mode matched and the product $\mathcal{M}=t_{1}t_{2}$ defines the degree of overlap.  

The probability $P_{n}$ of detecting $n$ photons is given by the convolution  of the joint probability distribution $P_{\textrm{\tiny{joint}}}(p,m,q)$. The parameters  $p$, $m$  and $q$  label the photon number in the three modes of the network: the signal ancilla,  the displaced signal, and the reference ancilla. 
Using the multimode theory of photo detection~\cite{L.Mandel1995}, we can simplify the evaluation of the convolution by expressing the photon number distribution $P_{n}$ in the form of Eq.~(\ref{P_N}), 
\begin{eqnarray}
 P_{n} &=& \braket{:\frac{  \hat{N}_{tot}  ^{n}}{n!} e^{-\hat{N}_{tot}}:}\,, 
 \end{eqnarray}
where the  photon number operator $\hat{N}_{tot}$ of the multimode field is equal to the sum of the number operators of each different detection mode
$\hat{N}_{tot}= \hat{p}^{\dagger} \hat{p}+ \hat{m}^{\dagger} \hat{m}+ \hat{q}^{\dagger} \hat{q}$ .  
Applying the beam splitter transformation of Eq.~(\ref{beamsplitter}), the evolution of the different modes  are given by
 \begin{equation}
 \hat{p} = \sqrt{T}(\sqrt{t_{1}}\hat{v}_{1}-\sqrt{1-t_{1}}\hat{a}) + \sqrt{1-T}\hat{v}_{3},
 \end{equation}
  \begin{equation}
 \hat{m} = \sqrt{T}(\sqrt{t_{1}}\hat{a}+\sqrt{1-t_{1}}\hat{v}_{1}) + \sqrt{1-T}(\sqrt{t_{2}}\hat{b}+\sqrt{1-t_{2}}\hat{v}_{2}), \quad \textrm{}
 \end{equation}
  \begin{equation}
 \hat{q} = \sqrt{T}\hat{v}_{4} + \sqrt{1-T}(\sqrt{t_{2}}\hat{v}_{2}-\sqrt{1-t_{2}}\hat{b})\,,
 \end{equation}
 in which $\hat{v}_i$, $i=1,2,3$, are auxiliary operators that represent the modes of the vacuum, as depicted in Fig.~\ref{3BS_overlap}.
Once more, considering coherent states, the operators can be replaced by complex numbers. The number operator $\hat{N}_{tot}$ can be substituted by
\begin{eqnarray}
\hat{N}_{tot} &\rightarrow& T \hat{a}^{\dagger}\hat{a} + (1-T) |\beta|^2 + \sqrt{T(1-T)\mathcal{M}}(\beta^*\hat{a}+\hat{a}^{\dagger}\beta) \nonumber \\
&\rightarrow& T\hat{D}^{\dagger}(\xi)\hat{a}^{\dagger} \hat{a} \hat{D}(\xi) + |\zeta|^{2}\,,
\end{eqnarray} 
where $\xi = \sqrt{\mathcal{M}}\sqrt{\frac{1-T}{T}}\beta$ is the amount of displacement. The parameter
$|\zeta| = \sqrt{(1-\mathcal{M})(1-T)}|\beta|$ can be interpreted as the amplitude of an effective coherent field. 

Finally, the probability of $n$ counts is given by
 \begin{eqnarray}
 P_{n}&=& \braket{:  \frac{  \left (T  \hat{D}^{\dagger}(\xi)\hat{a}^{\dagger} \hat{a} \hat{D}(\xi)  +|\zeta|^{2}  \right)^{n}   }{n!} e^{- (T\hat{D}^{\dagger}(\xi)\hat{a}^{\dagger} \hat{a} \hat{D}(\xi) +|\zeta|^{2} )}  :}\nonumber \\
 &=&\sum_{j =0}^{n} \braket{:  \frac{  \left (T  \hat{D}^{\dagger}(\xi)\hat{a}^{\dagger} \hat{a} \hat{D}(\xi)  \right)^{j}   }{j!} e^{- T\hat{D}^{\dagger}(\xi)\hat{a}^{\dagger} \hat{a} \hat{D}(\xi)  }  :} \frac{ |\zeta|^{2(n-j)}}{(n-j)!} e^{-|\zeta|^{2}}\,.
 \label{P_convoluted}
  \end{eqnarray}
From Eq.~(\ref{P_convoluted}), we recognise that the photon number distribution  is a convolution of two different photon number distributions. The first contribution corresponds to the statistics of an ideally displaced quantum signal, detected with efficiency $T$, and the second corresponds to a Poissonian distribution. As a consequence, the network shown in Fig.~\ref{3BS_overlap} and its counterpart depicted in  Fig.~\ref{simple_overlap}, which is conceptually simplified, exhibit equivalent photon number probabilities. 

Our derivations lead to the conclusion that the detected photon number distribution can always be expressed as a convolution of the displaced state distribution and a Poissonian distribution. As first mentioned in~\cite{K.Banaszek2002},  a Wigner function constructed with the distribution in Eq.~(\ref{P_convoluted}) is composed of the product of two separate Wigner functions. The decreasing overlap between the signal and reference fields distorts the reconstructed Wigner function such that it deviates towards a Gaussian multiplied with the value of the signal state Wigner function at the origin. This results in an artificial  broadening of the reconstructed Wigner functions of low-photon number Fock states.


\section{\label{diagnosis} Mode matching diagnosis}

\begin{figure}[!ht]
     \begin{center}
	\includegraphics[width  = 0.45\textwidth]{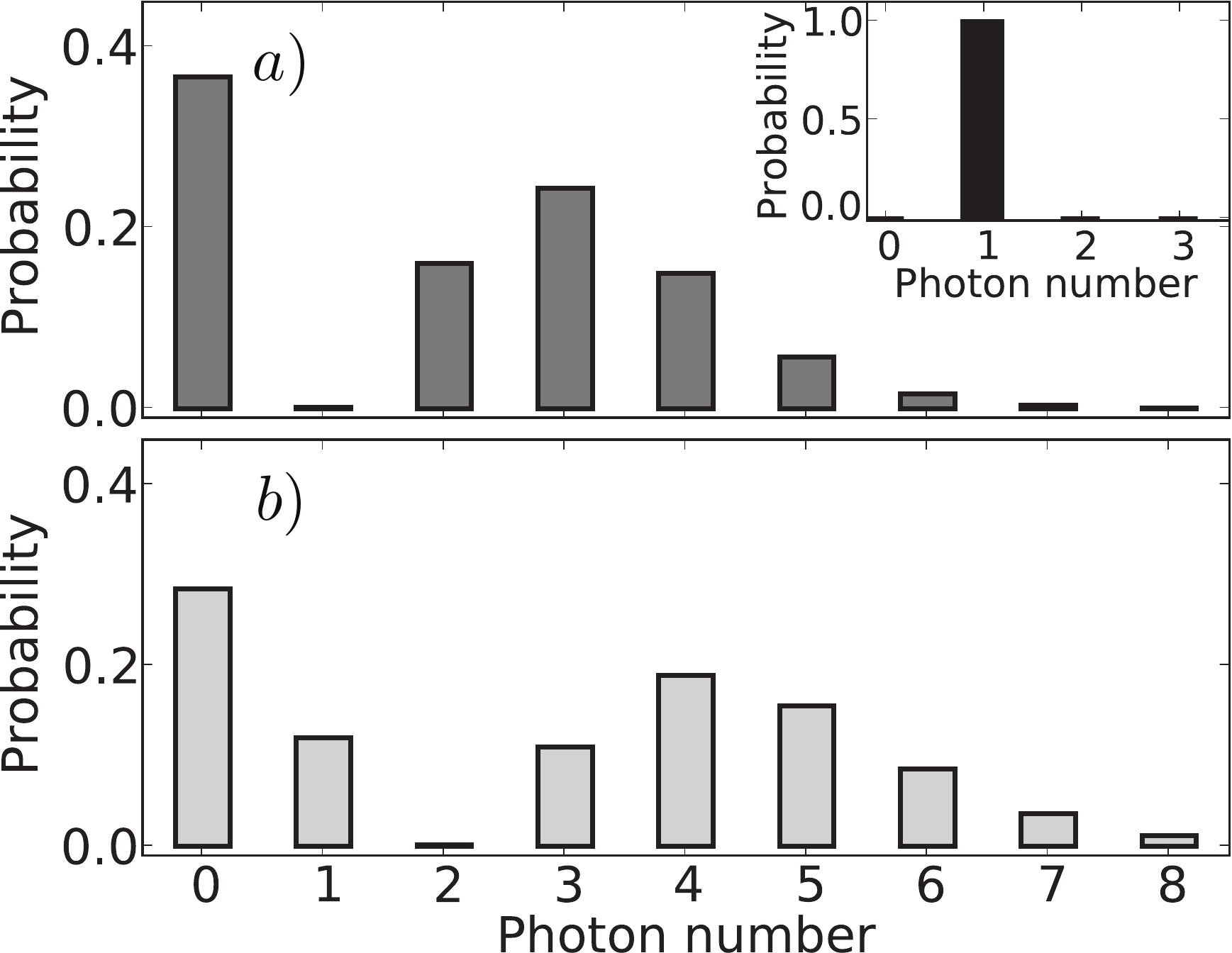}
	\includegraphics[width  = 0.45\textwidth]{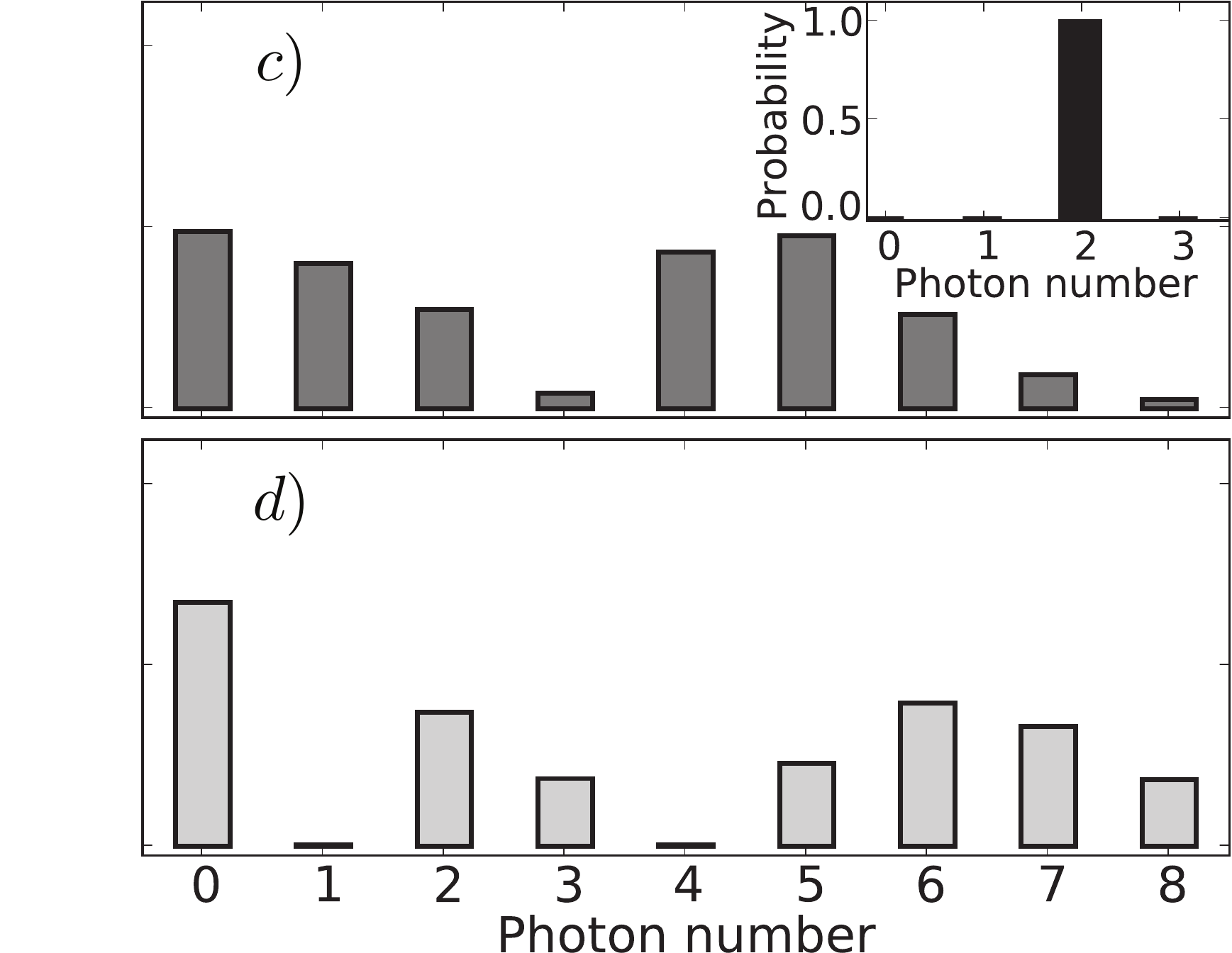}
	\caption{Photon number distribution for displaced a-b) single and c-d) two photon Fock states. The distributions are presented at two values of displacement:  $\alpha=1.0$  (dark grey) and $\alpha=1.4\sim \sqrt{2}$ (light grey). The insets show the photon number distributions at  $\alpha=0$ .}
	\label{95WignerP}
    \end{center}
\end{figure}

Beyond the investigation of loss-tolerant reconstruction of the Wigner function, direct probing also allows us to explore the non-trivial photon number distributions of displaced Fock states. As  a result of their exact photon number, Fock states exhibit a completely undetermined phase, which can be depicted by  a rotational symmetry of the field quadratures around the origin of the phase space. In other words, Fock states can be imagined as rings around the origin (uncertain phase) that have definite amplitudes (exact photon numbers). Photon number detection corresponds to evaluating the overlap of the studied signal state with Fock states. Using this concept it becomes intuitive that the photon number distributions of displaced Fock states must present oscillations as a consequence of an interference process in phase space \cite{W.Schleich1987, M.S.Kim1989, A.I.Lvovsky2002}. 
Fig.~\ref{95WignerP} presents typical  oscillations of the states $\ket{1}$  and $\ket{2}$. Considering e.g.~the displaced single photon Fock  state, we clearly see that the probability of  measuring one photon decreases as the displacement is increased. It is maximally suppressed when the displacement reaches the mean photon number $\braket{n_{\alpha}} = 1$, and increases again for higher values of the displacement. In this section this quantum behavior is explored in order to distinguish the effects of losses from those of imperfect mode overlap between the signal and reference fields.
 
\begin{figure}[!ht]
	\begin{center}
		\includegraphics[width  = 0.35\textwidth]{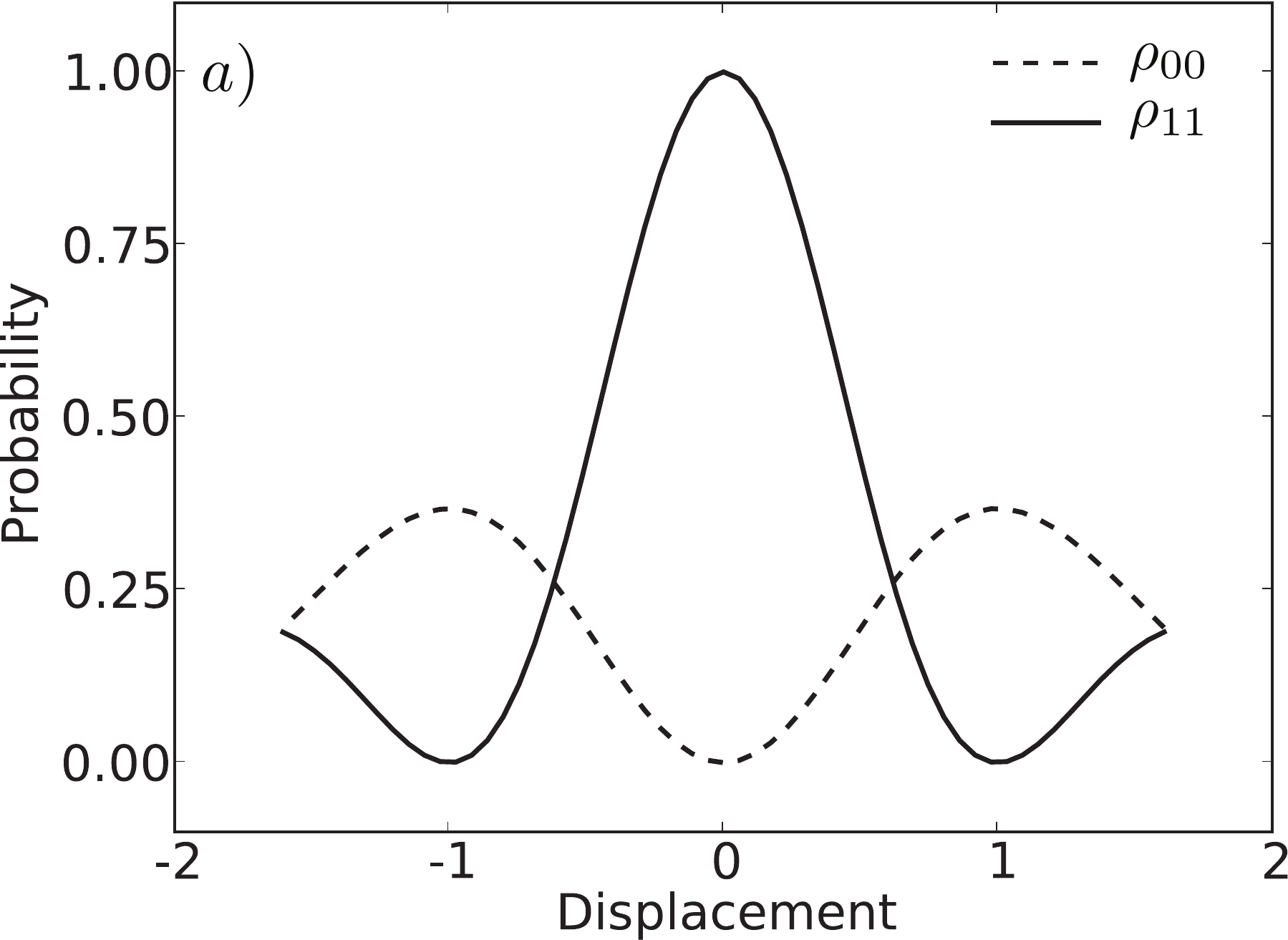}\hspace{-6mm}
		\includegraphics[width  = 0.35\textwidth]{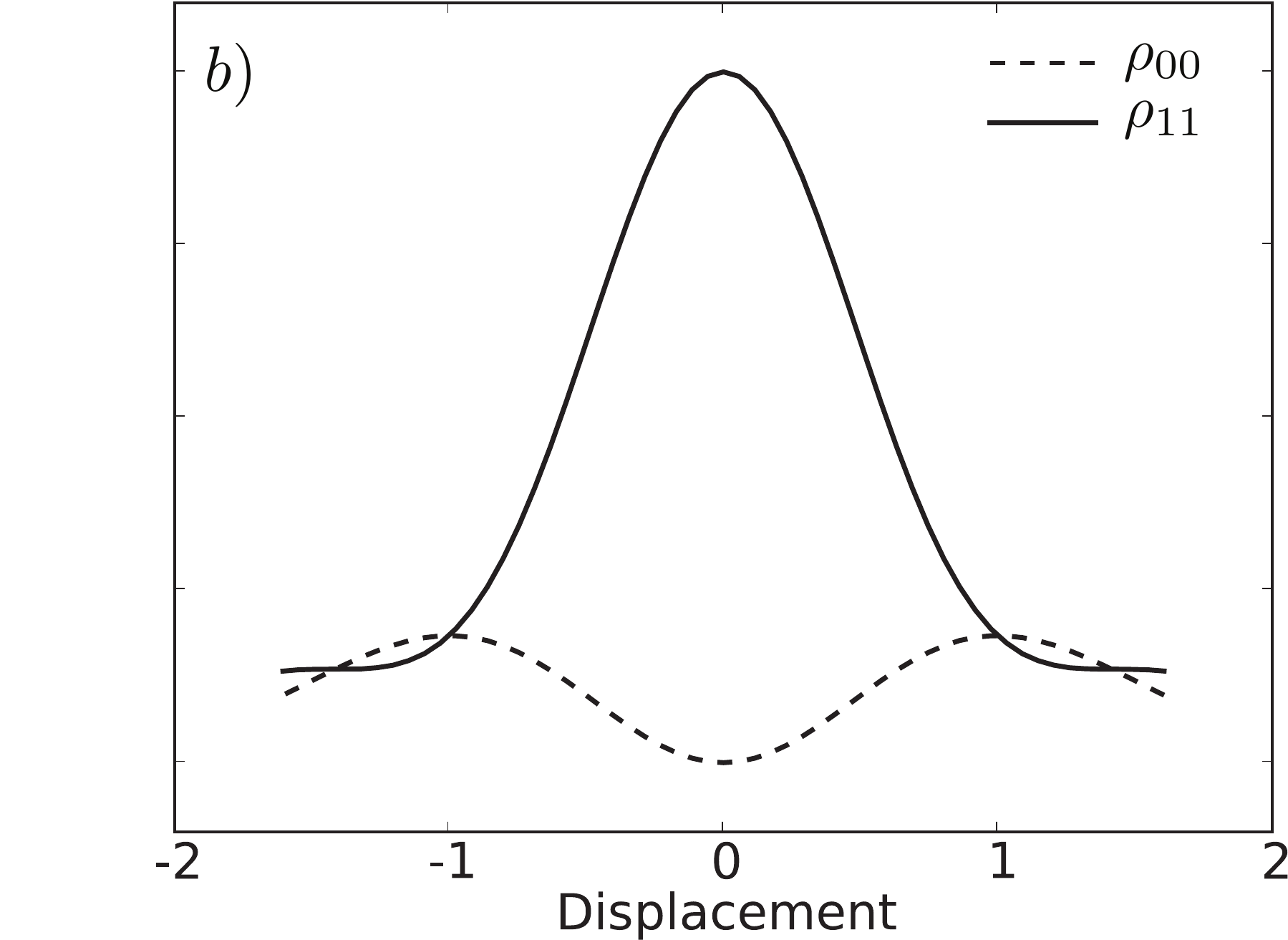}\hspace{-6mm}
		\includegraphics[width  = 0.35\textwidth]{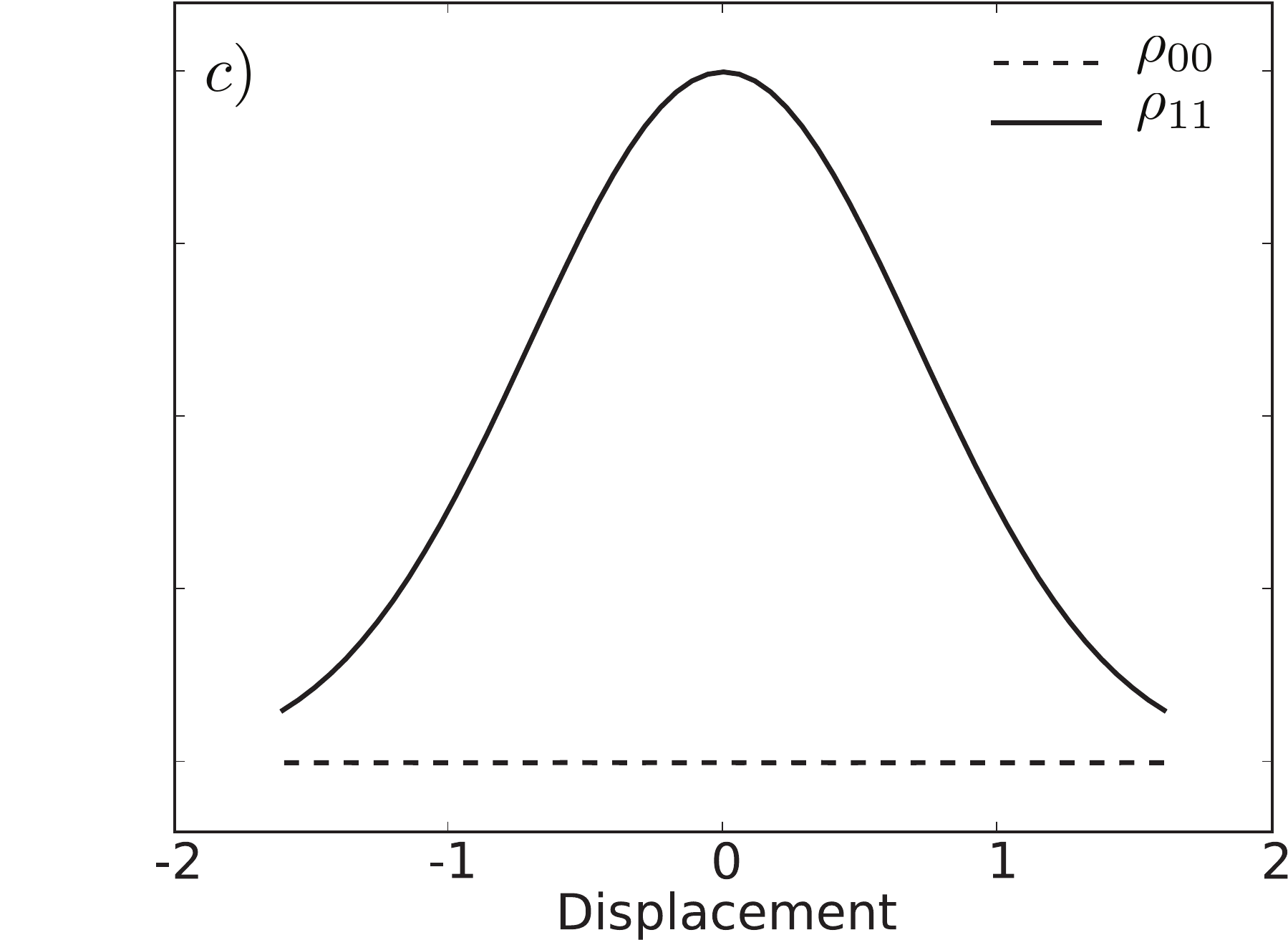}
		\caption{Behaviour of the vacuum and one photon components  of the statistics vs.~the amplitude of the displacement applied to  the single photon Fock state $\ket{1}$ in the cases of a) perfect overlap ($\mathcal{M} = 1$), b) 50\% overlap ($\mathcal{M}= 0.5$) and c) no overlap ($\mathcal{M} = 0$). }
		\label{overlaps}
	\end{center}
\end{figure}

We numerically study the reconstruction of the photon number distribution of the single photon Fock state as a function of the degree of overlap $\mathcal{M}$ introduced in the last section. 
The photon number distribution is evaluated using  Eq.~(\ref{P_convoluted}), which gives for  $P_{0}$ and  $P_{1}$   the following expressions
\begin{equation}
\hspace{-20pt}P_{0}= e^{-(1-T)|\beta|^{2}}(1-T) (1+T\mathcal{M}|\beta|^{2} ),
\label{overlapP0}
\end{equation}
\begin{equation}
\hspace{-20pt}P_{1}= e^{-(1-T)|\beta|^{2}}\left [ T + (1-T)|\beta|^{2}(1-T-2\mathcal{M}T) + (1-T)^{2}|\beta|^{4}  \mathcal{M}T\right].
\label{overlapP1}
\end{equation}
We consider a beam splitter with $T=0.95$ and reconstruct the photon number distributions for the cases of 100\%, 50\% and 0\% overlap applying the loss inversion algorithm and scaling of the displacement as discussed previously. 
Fig.~\ref{overlaps}  illustrates the characteristic behaviour of the vacuum and one photon components 
of the photon number distribution with respect to the displacement applied to single photon Fock state.
In the case of perfect overlap between signal and reference we expect a complete suppression of the one photon component when the displacement is $\alpha=1$ [Fig.~\ref{overlaps}(a)], for 50\% overlap, the oscillation of the vacuum and one photon components is reduced but still visible [Fig.~\ref{overlaps}(b)]. However, for non-overlapping fields the oscillation is completely washed out and the photon number distribution is just a convolution of the single photon state statistics with a Poissonian [Fig.~\ref{overlaps}(c)]. A clear signature of the lack of mode matching becomes apparent in the vacuum component of the inverted statistics: in the case of non-overlapping fields this contribution is always zero independent of the displacement. This effect is opposed to the signal degradation caused by losses, which always increases the probability of obtaining the zeroth photon number component.

\begin{figure}[!ht]
	\begin{center}
		\includegraphics[width  = 0.4\textwidth]{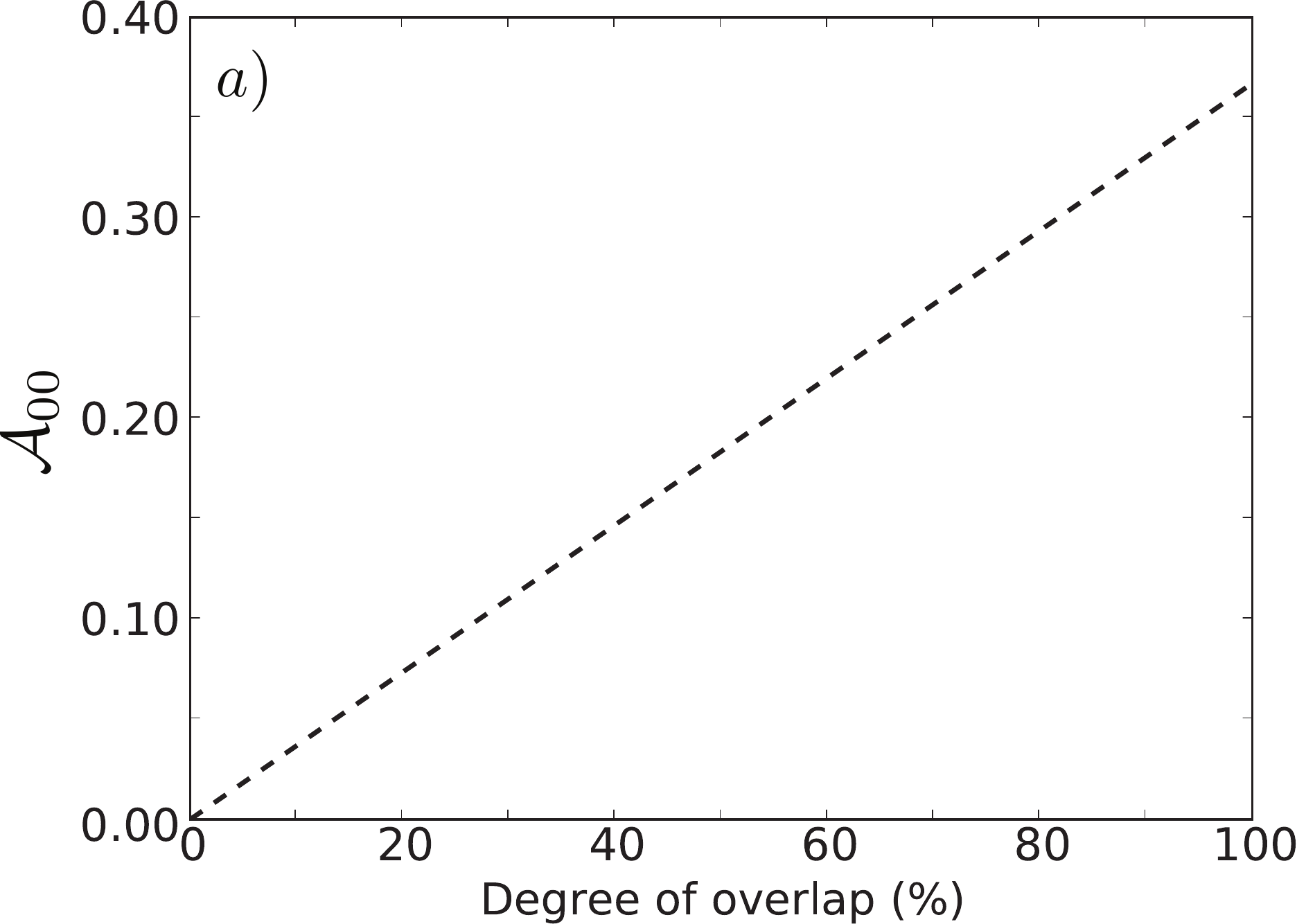}
		\includegraphics[width  = 0.4\textwidth]{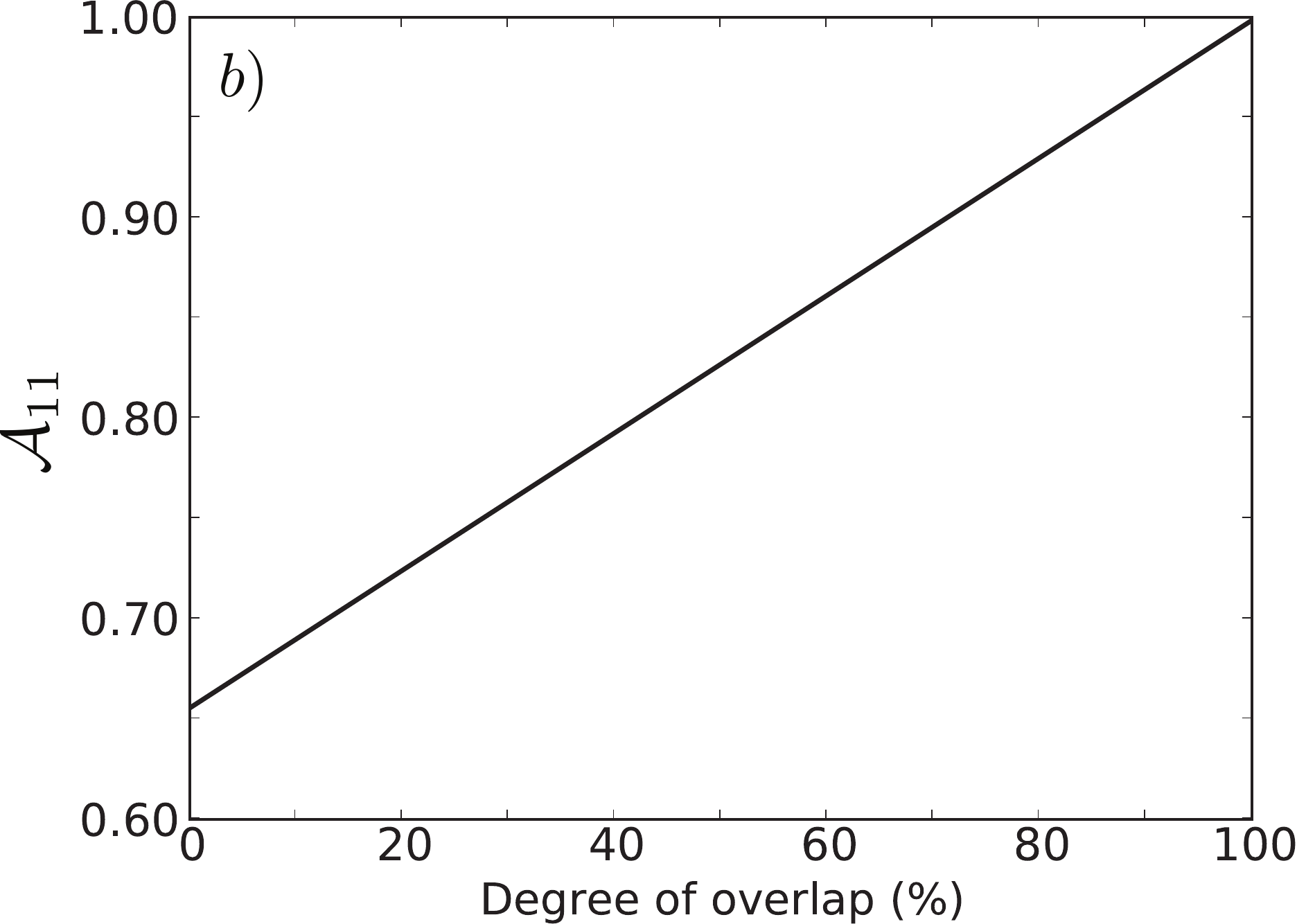}
		\caption{Amplitude of oscillation of the  a) vacuum component, $\mathcal{A}_{00}=|\rho_{00}(1)-\rho_{00}(0)|$ and b) one photon component, $\mathcal{A}_{11}=|\rho_{11}(1)-\rho_{11}(0)|$ with respect to the degree of overlap. A displaced single photon Fock state is considered. }
		\label{oscillation}
	\end{center}
\end{figure}

These results suggest a recipe to directly deduce the degree of mode matching from the amount of oscillation  of the vacuum and one photon components. In Fig.~\ref{oscillation} we show the values of $\mathcal{A}_{00}=|\rho_{00}(1)-\rho_{00}(0)|$ and $\mathcal{A}_{11}=|\rho_{11}(1)-\rho_{11}(0)|$ as a function of $\mathcal{M}$, where $\rho_{ii}(\alpha)$ is the probability of having $i$ photons in the photon number distribution at the displacement $\alpha$. 
Thus, $\mathcal{A}_{ii}$ represents a measure of the amplitude of the oscillation. From an experimental point of view, this method is an extremely direct way of characterizing the amount of overlap, as it solely  requires the measurement of the vacuum (or one photon) component without displacement and at a single point with $\alpha=1$.


\section{\label{highlosses} Monte Carlo Simulations}

In order to characterize a quantum state experimentally it has to be prepared several times enabling an ensemble measurement, which allows us to reconstruct the photon number statistics of the state. Here, a similar scenario was achieved using a Monte Carlo simulation. The detected click statistics were simulated with the help of Eqs.~(\ref{P_displaced1}) and (\ref{P_displaced2}), which describe the photon number distribution of displaced Fock states. We then replaced the parameter $T$ with the value of the overall loss $\eta$ and recalibrated the value of the displacement accordingly. By performing a Monte Carlo simulation we generated the statistics of the studied quantum states, which were then convolved with the known stochastic description of the TMD.
As the states under consideration ($\ket{1}$ and $\ket{2}$) are symmetric in phase space, we assumed fixed values for the relative phase between signal and reference beam  
and changed only  the amplitude of the displacement. 

To recover the photon number statistics of the displaced state we employed the direct numerical inversion of Eq.~(\ref{CL_inversion}) and evaluated the photon number parity from the statistics to yield the reconstructed value of the Wigner function. 
In order to estimate the statistical errors, for each sampled point of the phase space  we performed ten Monte Carlo simulations, each consisting of $10^{6}$ events. The chosen number of events was based on the expected number of experimental trials; this depends, for example, on the setup stability. This set of simulations gave at each sampled point ten simulated values for the Wigner function, from which we evaluated the mean value and the standard deviation.

There are two factors limiting the range over which the state reconstruction can be considered reliable. First,  as  a consequence of the TMD resolution,
the mean photon number of the displaced beam is restricted. With a $m$-stage TMD (incorporating $m$ 50/50 couplers) we are able to resolve at maximum $2^{m}$ clicks (equal to the number of bins into which the incident pulse is separated). The reconstruction becomes unstable as soon as the displaced state has a significant  contribution of  photon numbers larger than the number of the TMD bins. In our numerical investigations, we observed that the instability can be tolerated until  the probability of  having more than $2^{m}$ photons in the displaced state is on the order of $2-3\%$. 
The second limitation arises from statistical errors and can be estimated from the amount of collected events versus losses. Statistical errors are highly amplified in the loss inversion process. As the losses increase the inversion becomes unstable,  yielding a singular loss matrix and negative probability components for the reconstructed photon number statistics. 
We define the reliable reconstruction range by a boundary  value for the displacement at which the inversion becomes unstable or, similarly, we accept the values of displacement at which the photon number statistics reconstructed from all $10^{7}$ events  render components $\rho_{nn} > -0.001$.

Any experimental implementation of the direct probing scheme, as envisioned with the current technology, is subjected to high losses. For example, the quantum efficiencies of APDs can reach values between 60-80\%. The coupling efficiency into single-mode fibres is highly mode-sensitive and can drop down to 30-40\%  in case of mismatched modes. Another important source of  degradation of the signal is the beam splitter used as displacing element. Typically, the transmittance of this beam splitter is around 95\%. 
Due to  the 5\% reflectance, the beam splitter intrinsically attenuates the quantum signal and modifies the displacement. Effectively, in a  realistic experiment for state characterization the achievable detection efficiency will lie between  20\%  and 30\%. In addition to the effects of inefficiencies or losses, we must take into account the accuracy with which the convolution matrix of the TMD can be determined. This is caused by unbalanced 50/50 couplers, which are generally specified with $5\%$  uncertainty.

 \subsection{\label{results}Results}
 We  performed the simulation for four different scenarios to clarify the effects of  different experimental frames.  We analysed the performance of the state characterization and estimated the limitations by considering a realisable displacement operator  implemented with a beam splitter of 95\% transmittance, a detection with three-stage (8-bin) TMD and perfect efficiency. Furthermore, we analysed the possible benefits associated with the use of a four-stage (16-bin) TMD taking also into account a finite detection efficiency. Next, the amount of degradation was increased, considering the experimentally achievable efficiency level. Finally, the effect of imperfect mode overlap was included.
 
\begin{figure}[!ht]
	\includegraphics[width  = 0.45\textwidth]{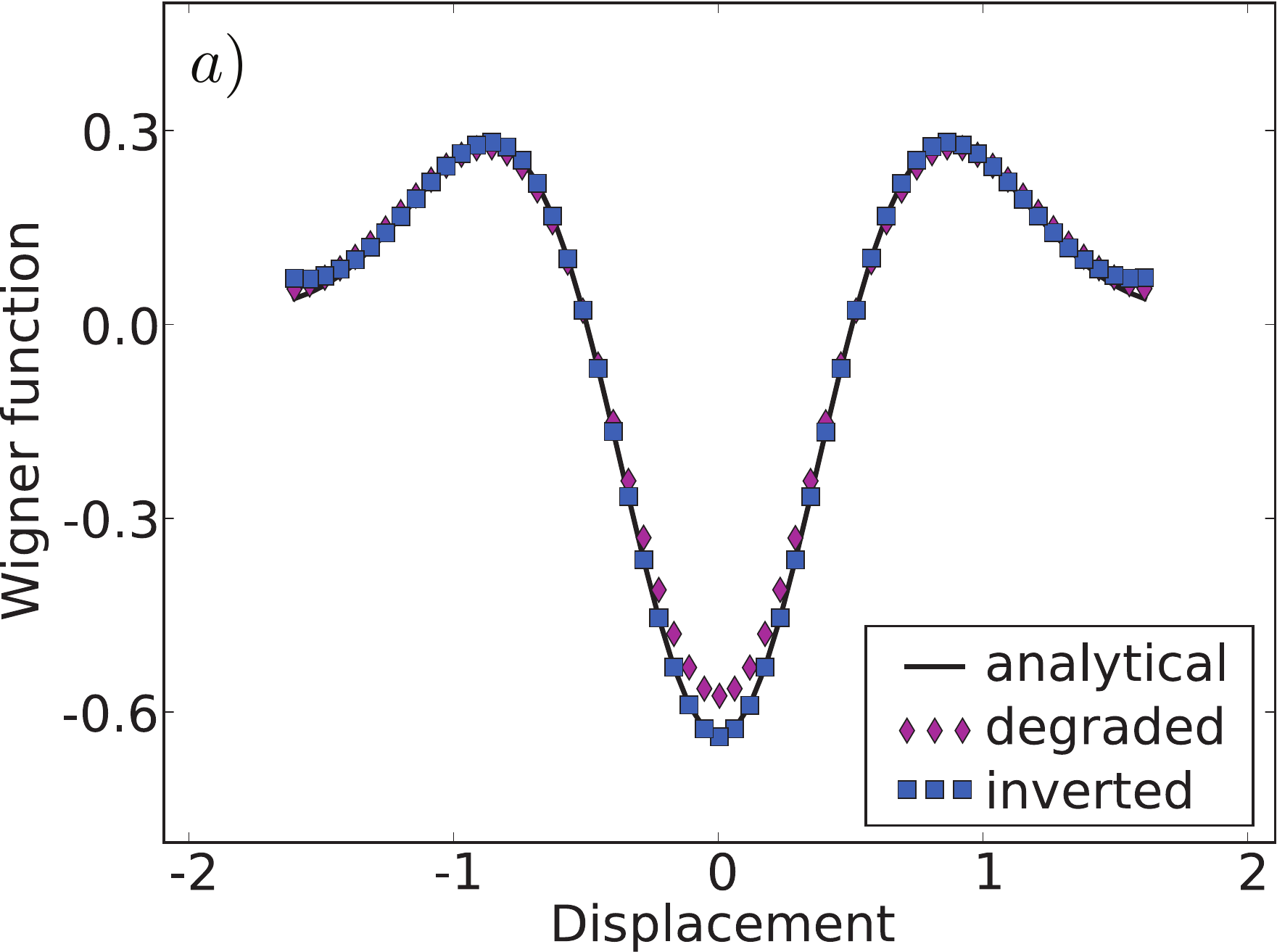}\hspace{10mm}
	\includegraphics[width  = 0.45\textwidth]{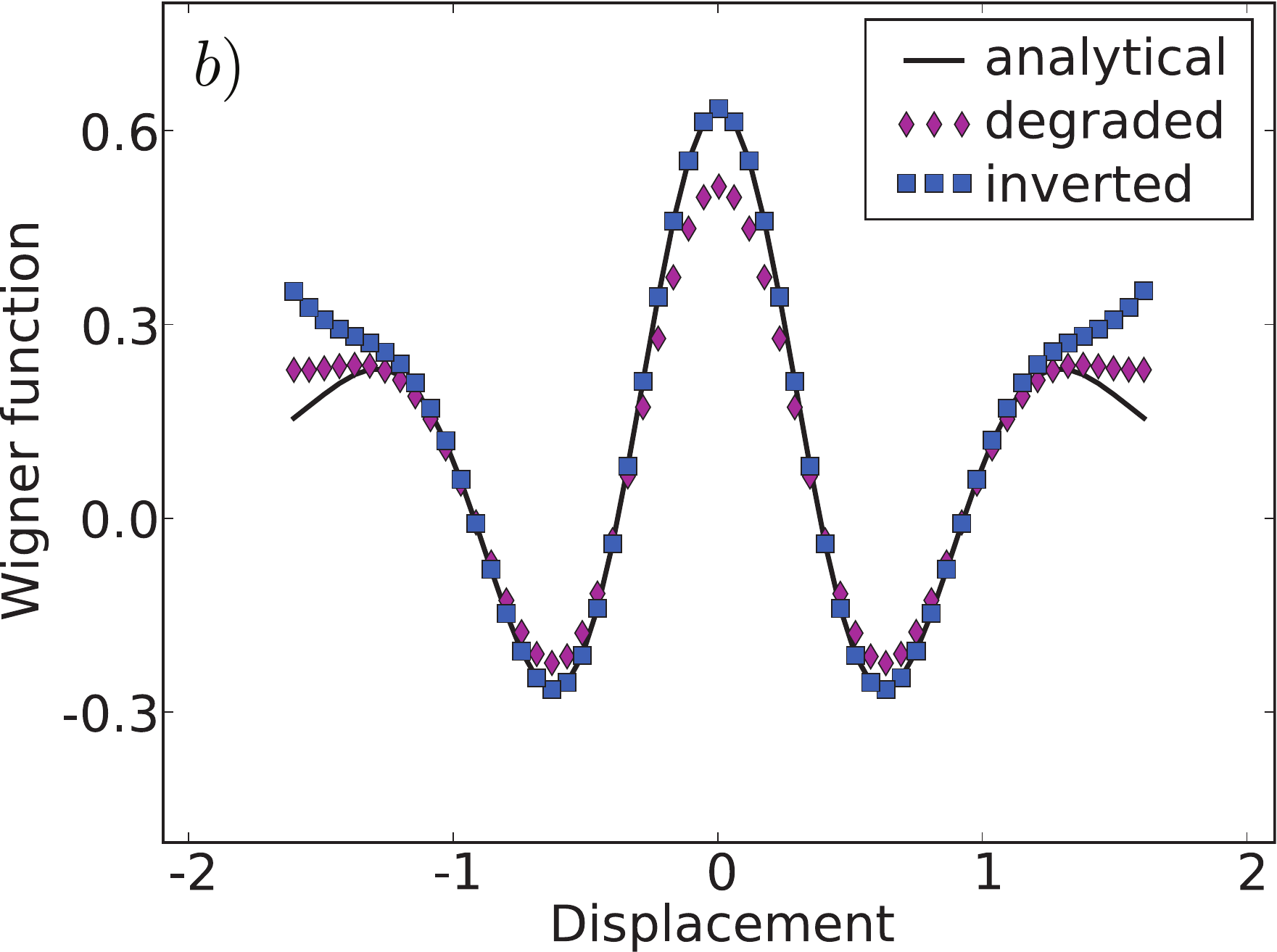}
	\caption{Loss-degraded (diamonds), reconstructed (squares), and analytical (solid line) Wigner functions for Fock states a) $\ket{1}$  and b) $\ket{2}$. The simulated setup has a displacing beam splitter with 95\% transmittance and ideal detection efficiency. The reconstruction is very accurate with errors smaller than used symbols. For further details see text.}
	\label{95Wigner}
\end{figure}

\paragraph{Case 1:}
First we studied the idealized case of perfect detection efficiency but included the finite resolution of a TMD due to its limited number of stages. The quantum signal was displaced at a beam splitter with 95\% transmittance and the TMD was assumed to be constructed from ideal 50/50 couplers. The results show that the reconstructed Wigner functions of the  $\ket{1}$ and $\ket{2}$ Fock states are slightly loss-degraded  around the origin due to the reflection of the signal at the beam splitter. Nevertheless, this degradation can be easily  handled with the direct inversion, as shown in Fig.~\ref{95Wigner}. Due to the limited number of TMD stages, a reliable reconstruction of $W(\alpha)$ is possible only inside a given scanning range of phase space, i.e. there is a maximum value of $\alpha$ that can be safely used. The ultimate boundaries for the reliable scanning ranges correspond to  displacements of $\alpha=1.5$ and $\alpha=1.2$ for states $\ket{1}$ and $\ket{2}$, respectively. Nevertheless, with lossless detection, the 8-bin TMD is adequate to scan the interesting regions of the phase space for both of these states. 

\paragraph{Case 2: }
In the second simulation we considered more realistic parameters, which might be accessible in the near future. We assumed again a  beam splitter with  95\% transmittance, and a detection efficiency of $60\%$. In this case we compared the performance of 8- and 16-bin TMDs for reconstructing the Wigner function of the single photon Fock state.  As an example of an unbalanced TMD we chose couplers with splitting ratios of 45/55. However, the deconvolution was still done by assuming perfect 50/50 couplers, as the imbalance of $5\%$ is due to uncertainty in the beam splitter specification. 
\begin{figure}[!ht]
	\includegraphics[width  = 0.45\textwidth]{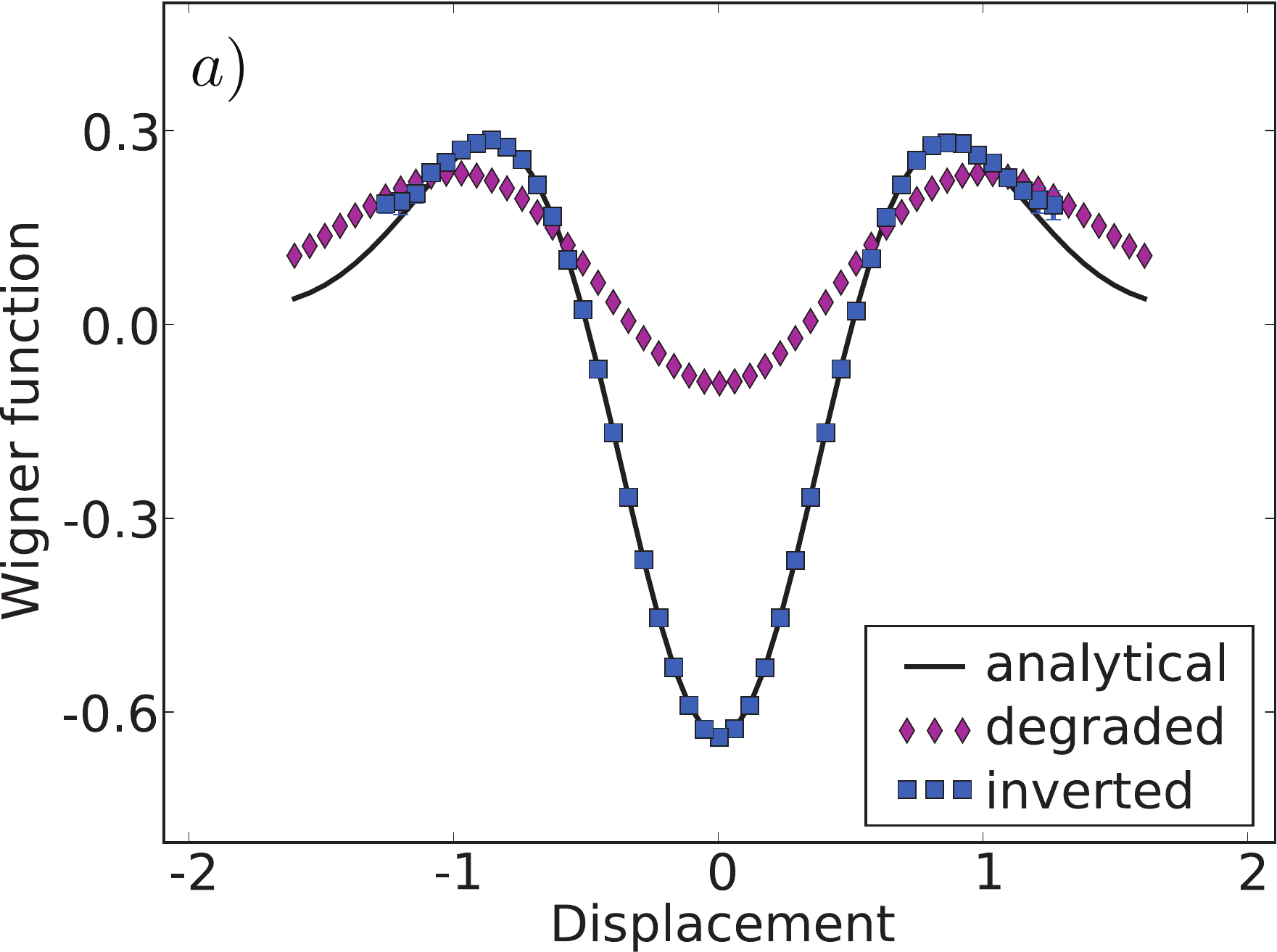}\hspace{10mm}
	\includegraphics[width  = 0.45\textwidth]{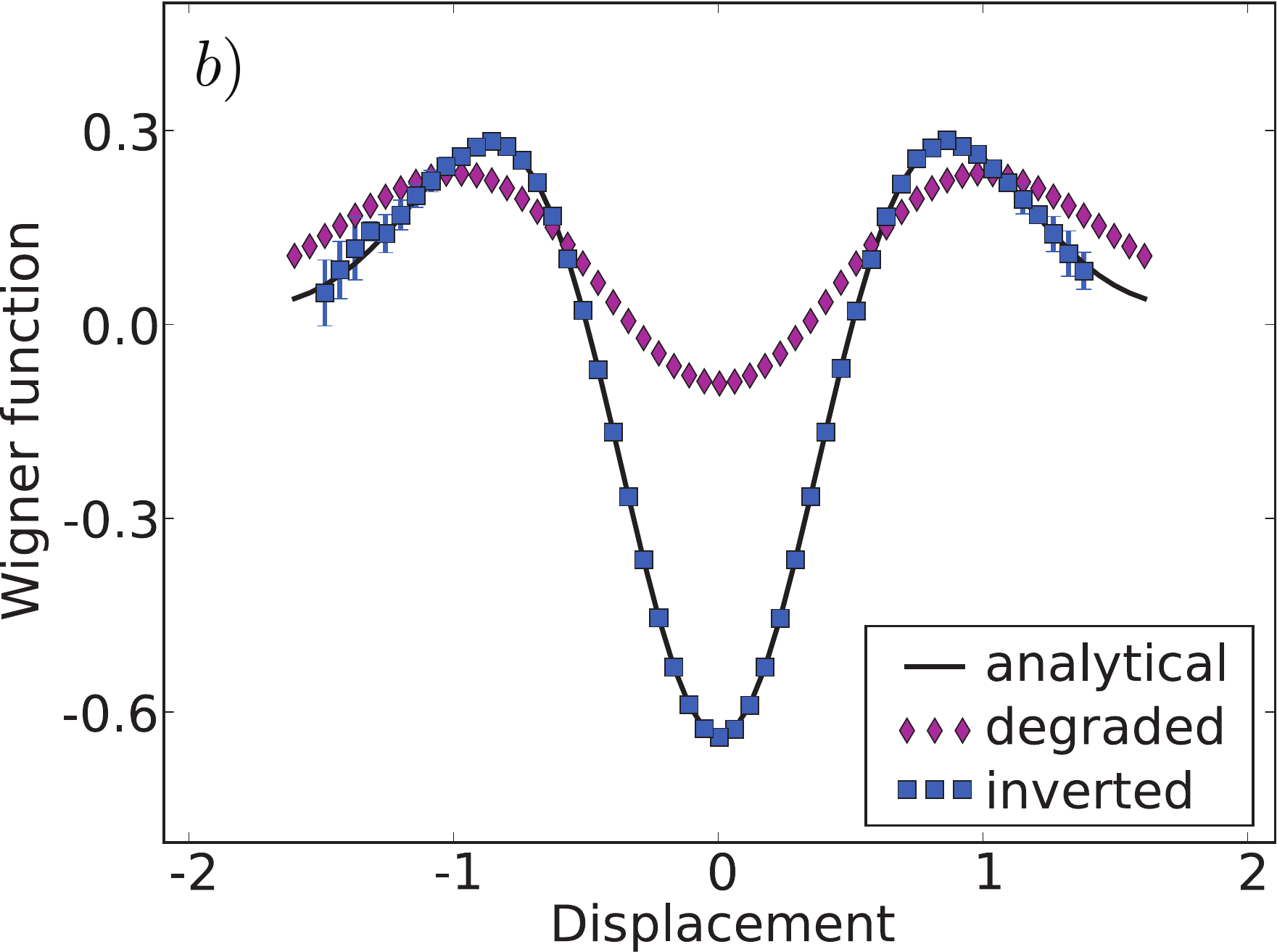}
	\caption{Loss-degraded (diamonds), reconstructed (squares), and analytical (solid line) Wigner functions for the single Fock state $\ket{1}$. a) 8-bin and b) 16-bin TMDs were analysed in a setup of  60\% detection efficiency. The reconstruction around the origin is very accurate with errorbars are smaller than used symbols.}
	\label{60Wigner}
\end{figure}
As the detection efficiency decreases, the effect of the loss becomes prominent restricting the reliable scanning range in phase space. Moreover, the effect of the unbalanced deconvolution becomes negligible in comparison to the effect of loss.   An inspection of Fig.~\ref{60Wigner} shows that very little advantage is gained in this loss regime by choosing the larger TMD. In this case the  direct inversion of the Fock state $\ket{1}$ breaks down when the displacement is close to $1.2$ 
 and $1.4$ for 8- and 16-bin TMDs, respectively.  

\begin{figure}[!ht]
	\includegraphics[width  = 0.45\textwidth]{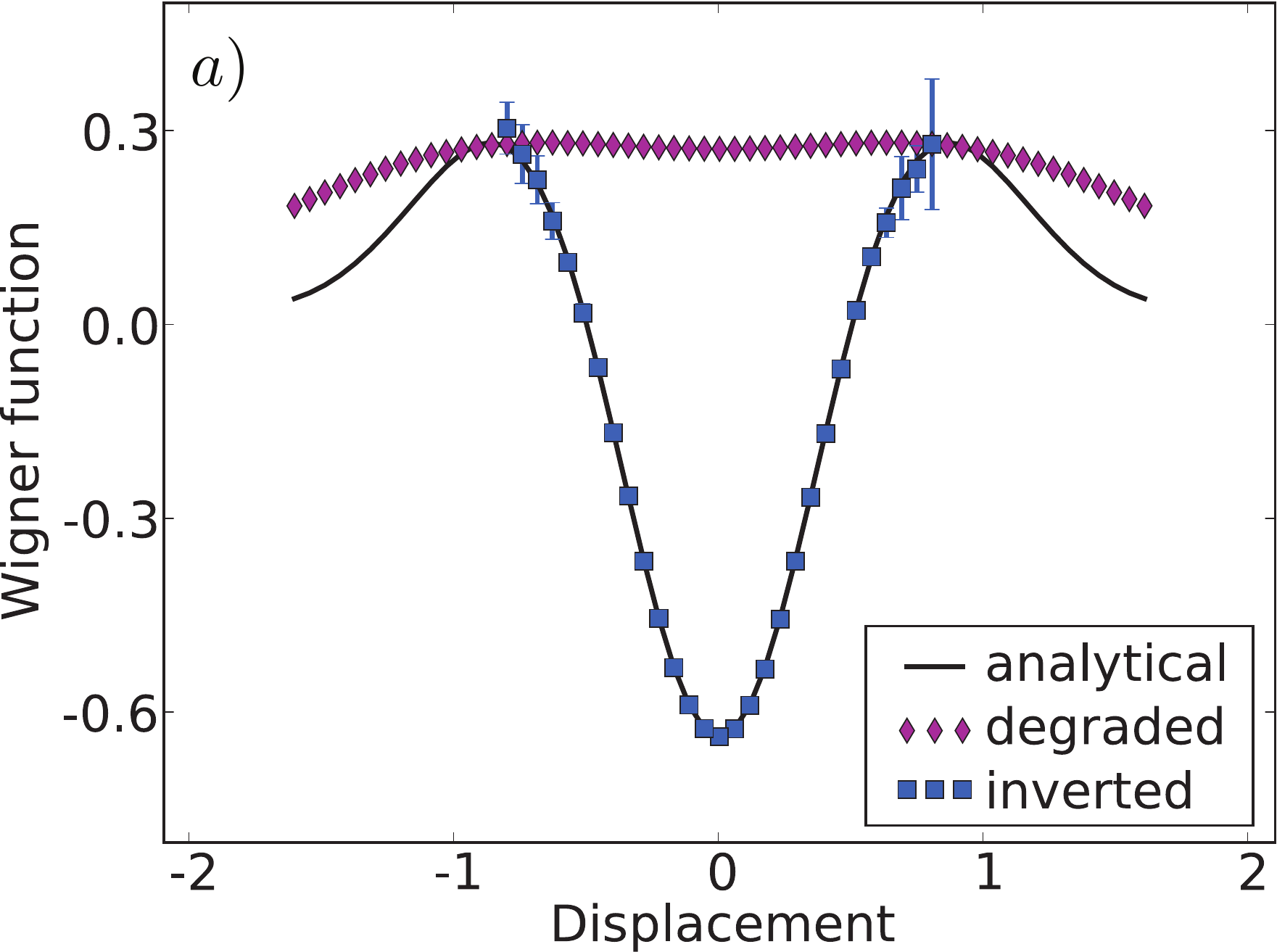}\hspace{10mm}
	\includegraphics[width  = 0.45\textwidth]{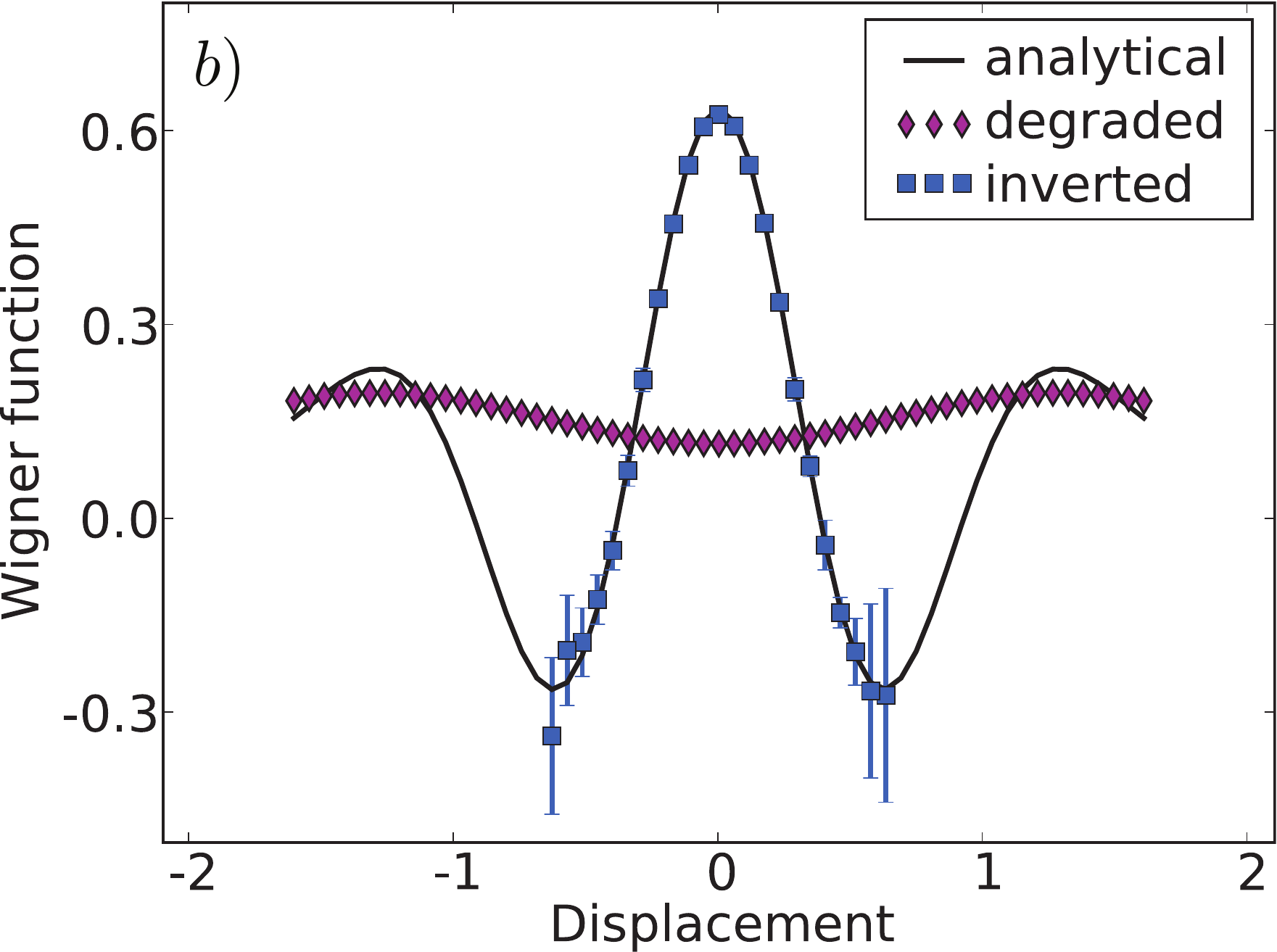}
	\caption{Loss-degraded (diamonds), reconstructed (squares), and analytical (solid line) Wigner functions for Fock states a) $\ket{1}$ and b) $\ket{2}$. In both simulations an 8-bin TMD and a detection efficiency  of $30\%$ was considered.}
	\label{30Wigner}
\end{figure}

\paragraph{Case 3: }
In the third simulation, we studied the reconstruction ranges for the single and two photon Fock states, assuming a currently feasible experimental situation in which the detection efficiency equals 30\%. Once again, the scheme consists of a beam splitter with 95\% transmittance, unbalanced TMD with 45/55 couplers but balanced deconvolution. At this high loss regime we can  choose the 
 8-bin TMD for photon counting, since the ability to record higher photon numbers does not bring any advantage to the loss inversion. In this situation,  the negative parts of the loss-degraded Wigner functions are completely washed out in the loss-degraded data. Nevertheless, the inner regions of the Wigner functions can still be reliably reconstructed up to displacements of 0.8 and 0.6 
for the states $\ket{1}$ and $\ket{2}$, respectively (Fig.~\ref{30Wigner}). In the case of the Fock state $\ket{2}$, the inversion causes negative probability components also close to the origin,  indicating that the inversion is unstable. This is not caused by the contribution of higher photon numbers but by the accuracy at which the vacuum, one and two photon components are recorded.  Nevertheless, the results appear to be accurate as the  error in the parity measurement is seen to be negligible.

\begin{figure}[!ht]
    \begin{center}
	\includegraphics[width  = 0.45\textwidth]{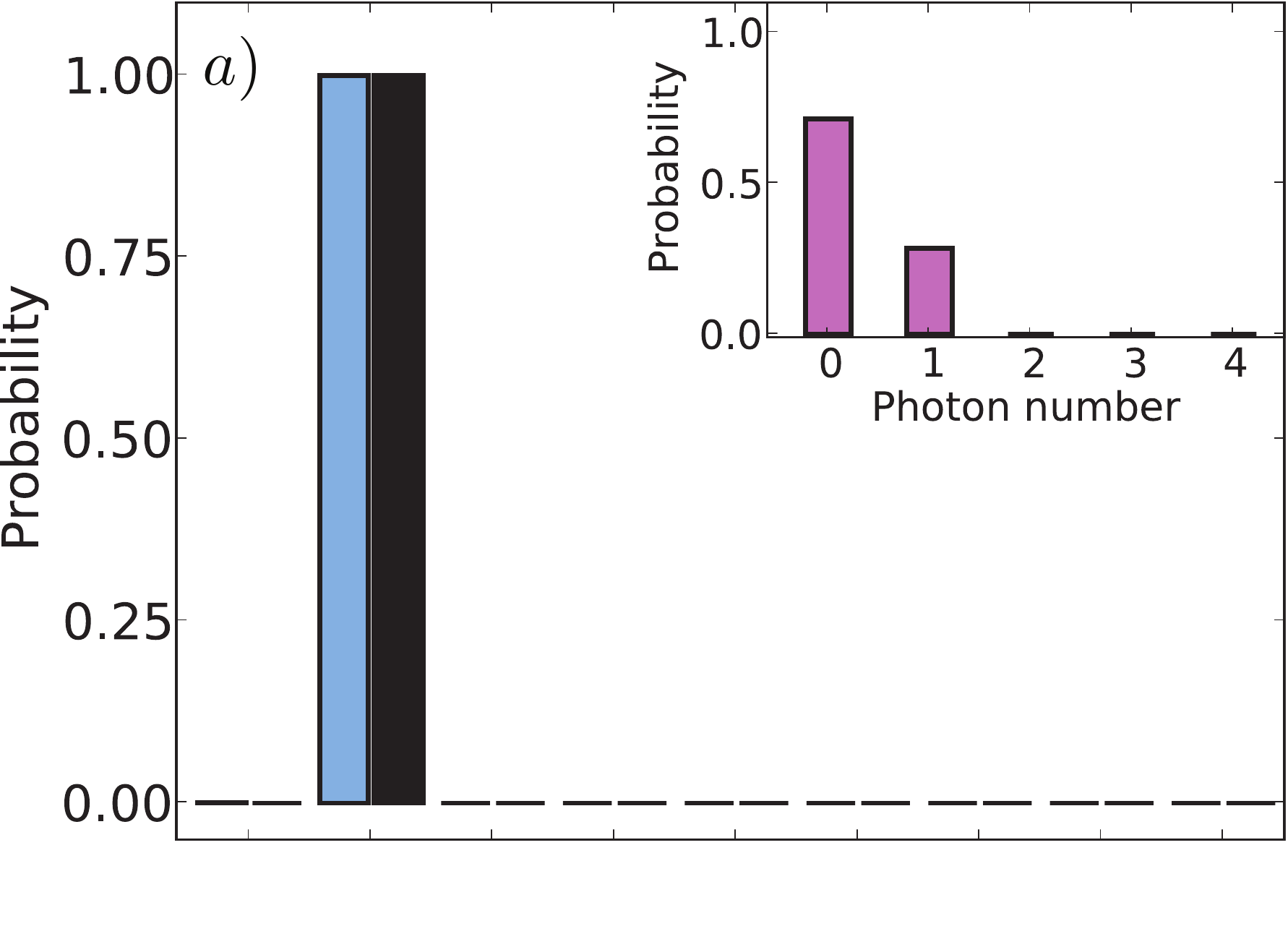}\hspace{10mm}
	\includegraphics[width  = 0.45\textwidth]{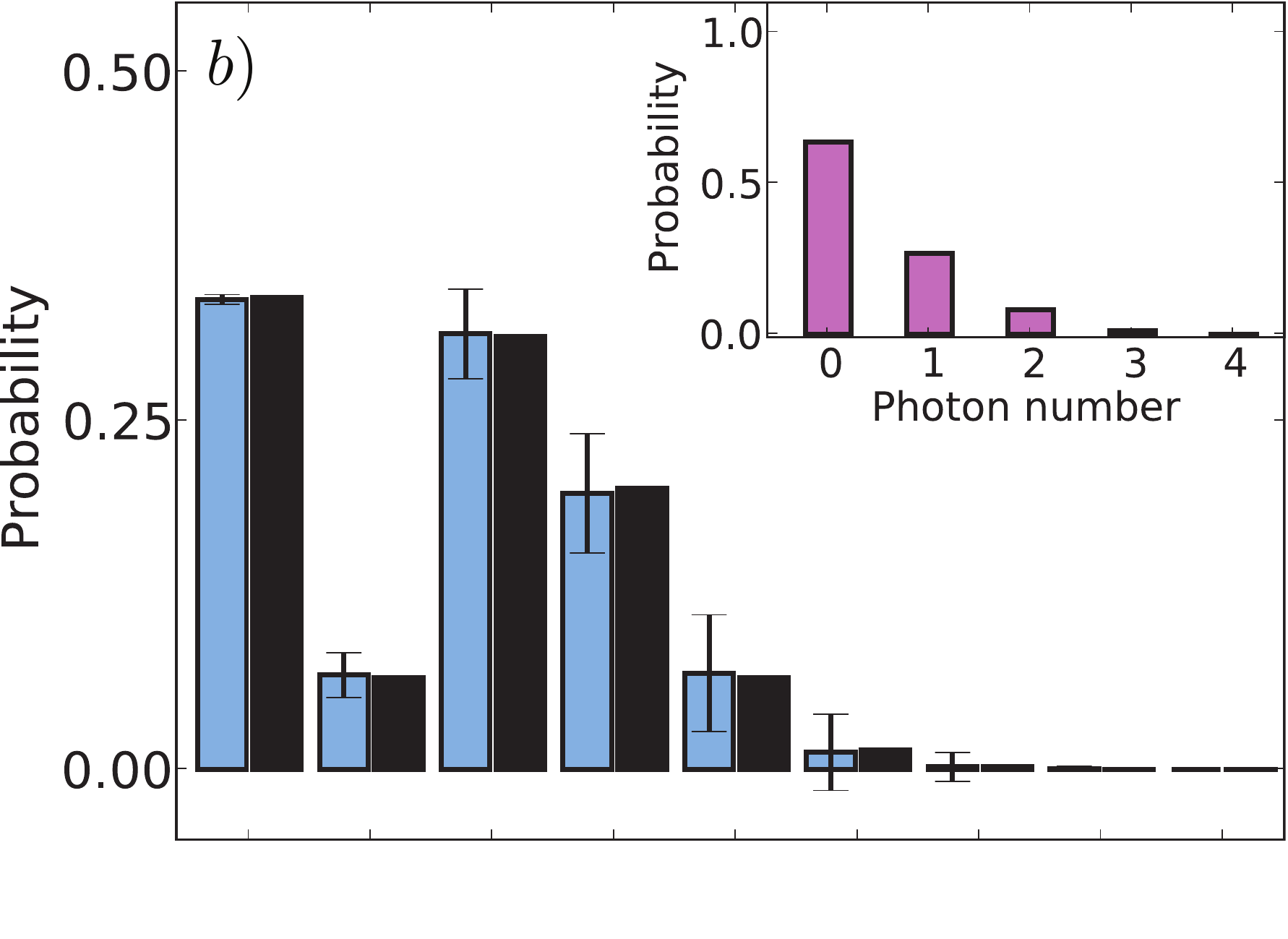}
	\includegraphics[width  = 0.45\textwidth]{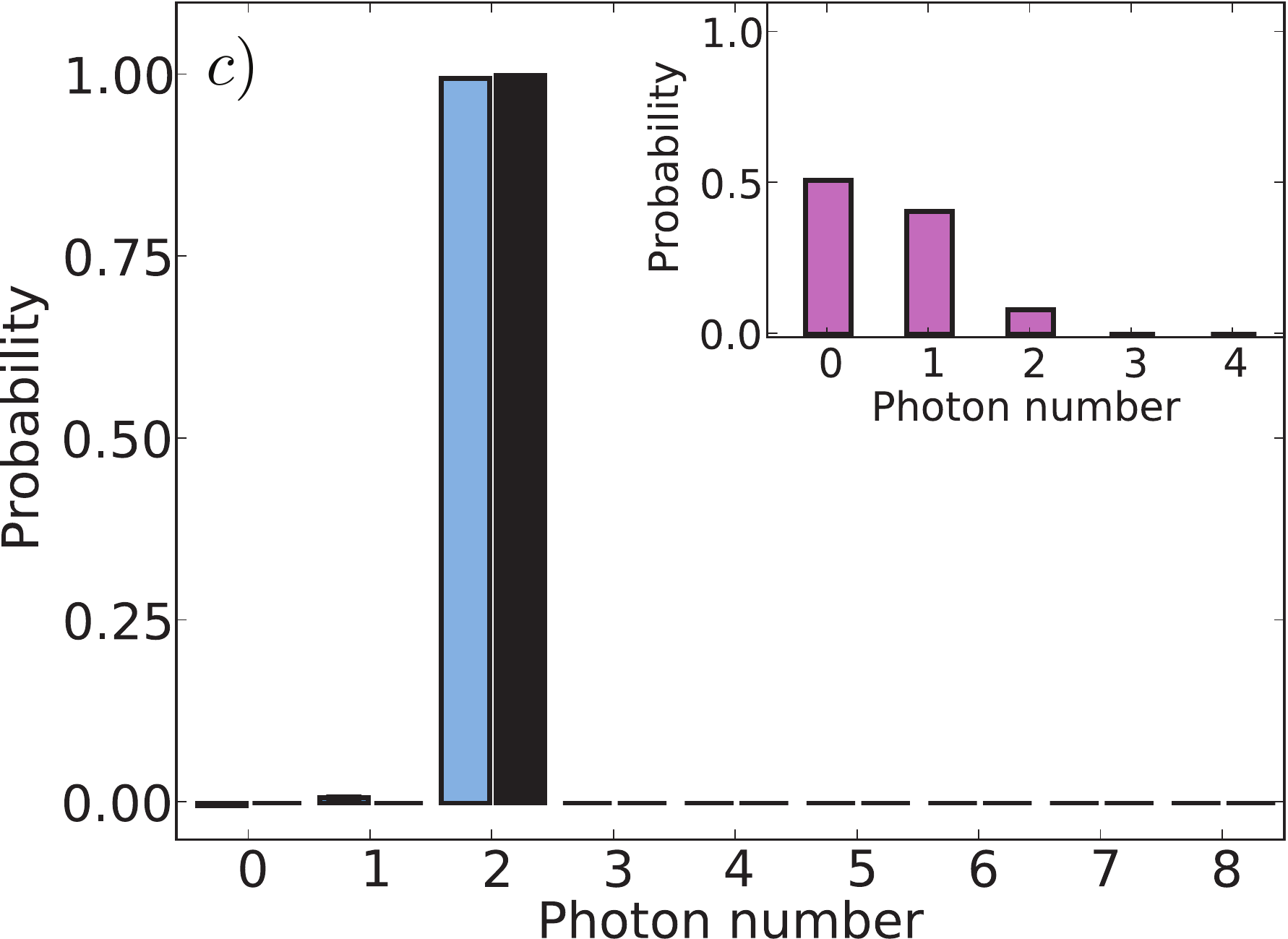}\hspace{10mm}
	\includegraphics[width  = 0.45\textwidth]{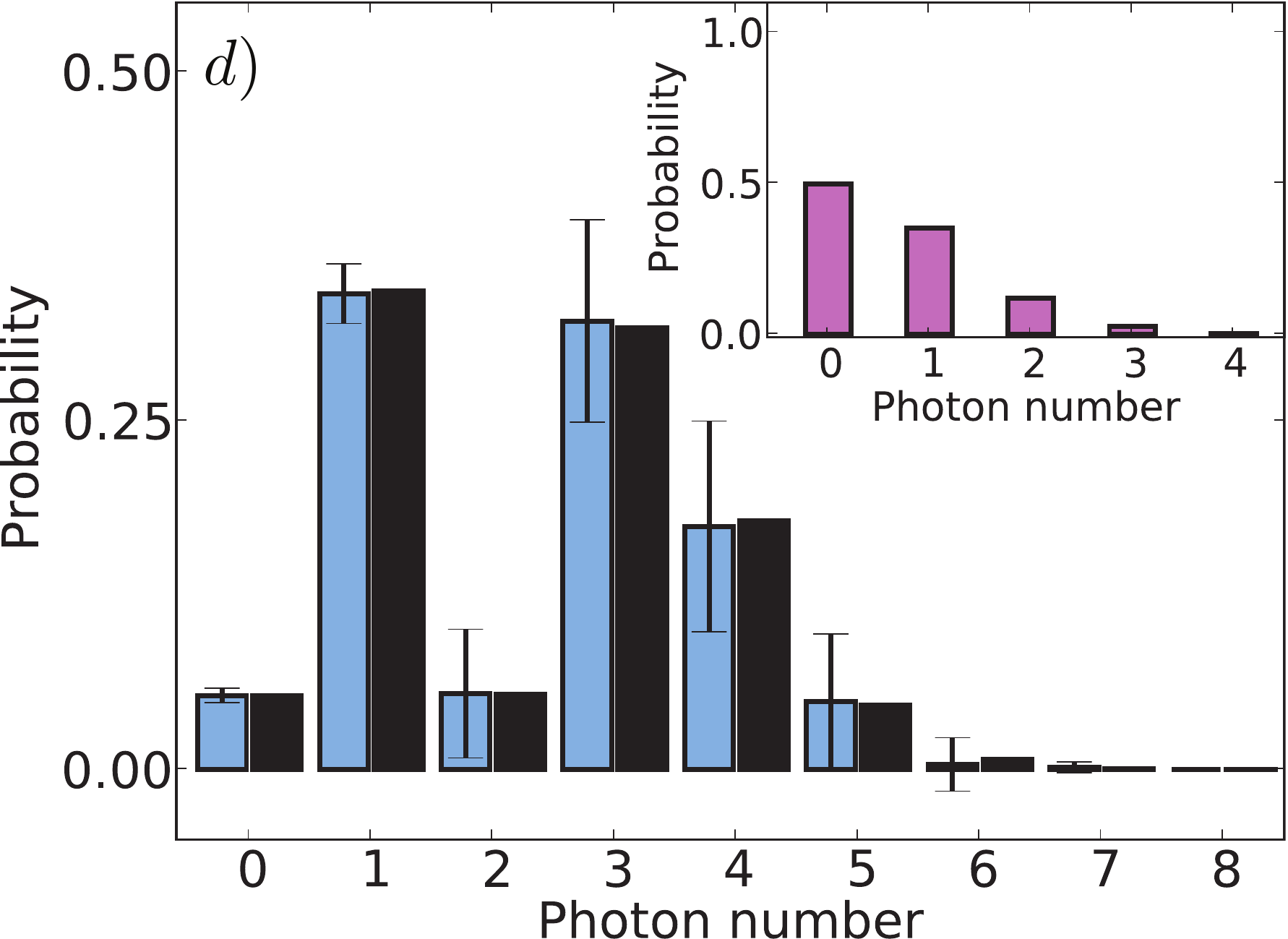}
	\caption{Simulated statistics  at the values of a) $\alpha = 0$ and b) $\alpha = 0.8$  for the single photon Fock state and at c) $\alpha = 0$  and d) $\alpha = 0.6$ for the two photon Fock state. The loss-degraded (pink, inset) and inverted (blue) statistics are shown together with the analytical solution of the photon number distribution (black).  The two extreme cases of no displacement (left side) and the highest displacement (right side) allows a  reliable state reconstruction despite of the low detection efficiency -- 30\%.}
	\label{30Probability}
	\end{center}
\end{figure}

Despite the low efficiency (30\%), the simulation indicates that we are able to reconstruct the oscillatory behaviour of the photon number statistics for the Fock states $\ket{1}$ and $\ket{2}$. The loss obscures the oscillations of the photon statistics, but similarly to the Wigner function analysis, the inversion technique enables us to reconstruct this characteristic behavior from the TMD measurement. In Fig.~\ref{30Probability} we show this behavior,  comparing  the loss-degraded,  inverted, and expected photon statistics at zero displacement and at  the highest reliable displacement that can be applied to the quantum state.

\begin{figure}[!ht]
	\begin{center}
		\includegraphics[width  = 0.5\textwidth]{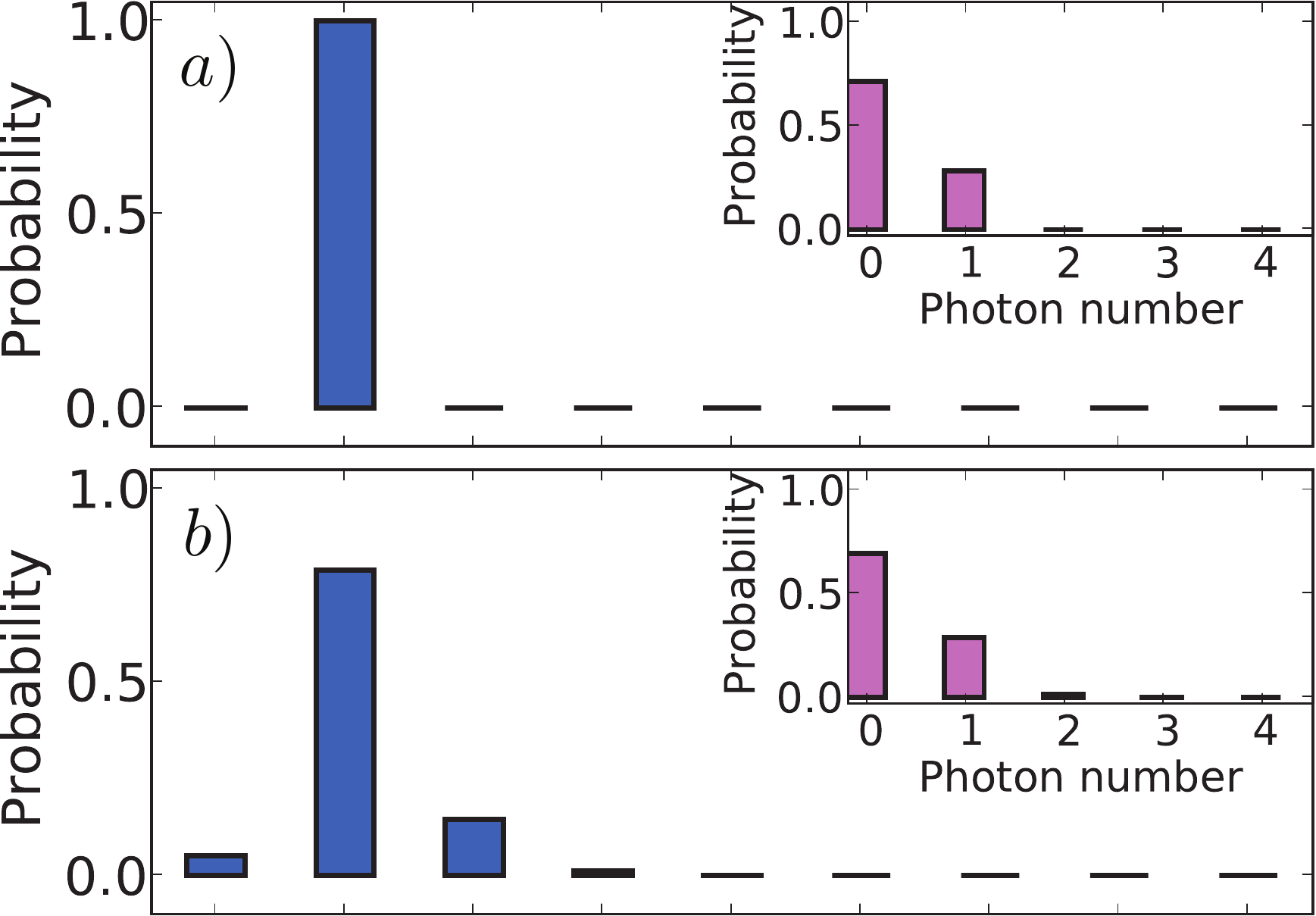}\vspace{1mm}
		\includegraphics[width  = 0.5\textwidth]{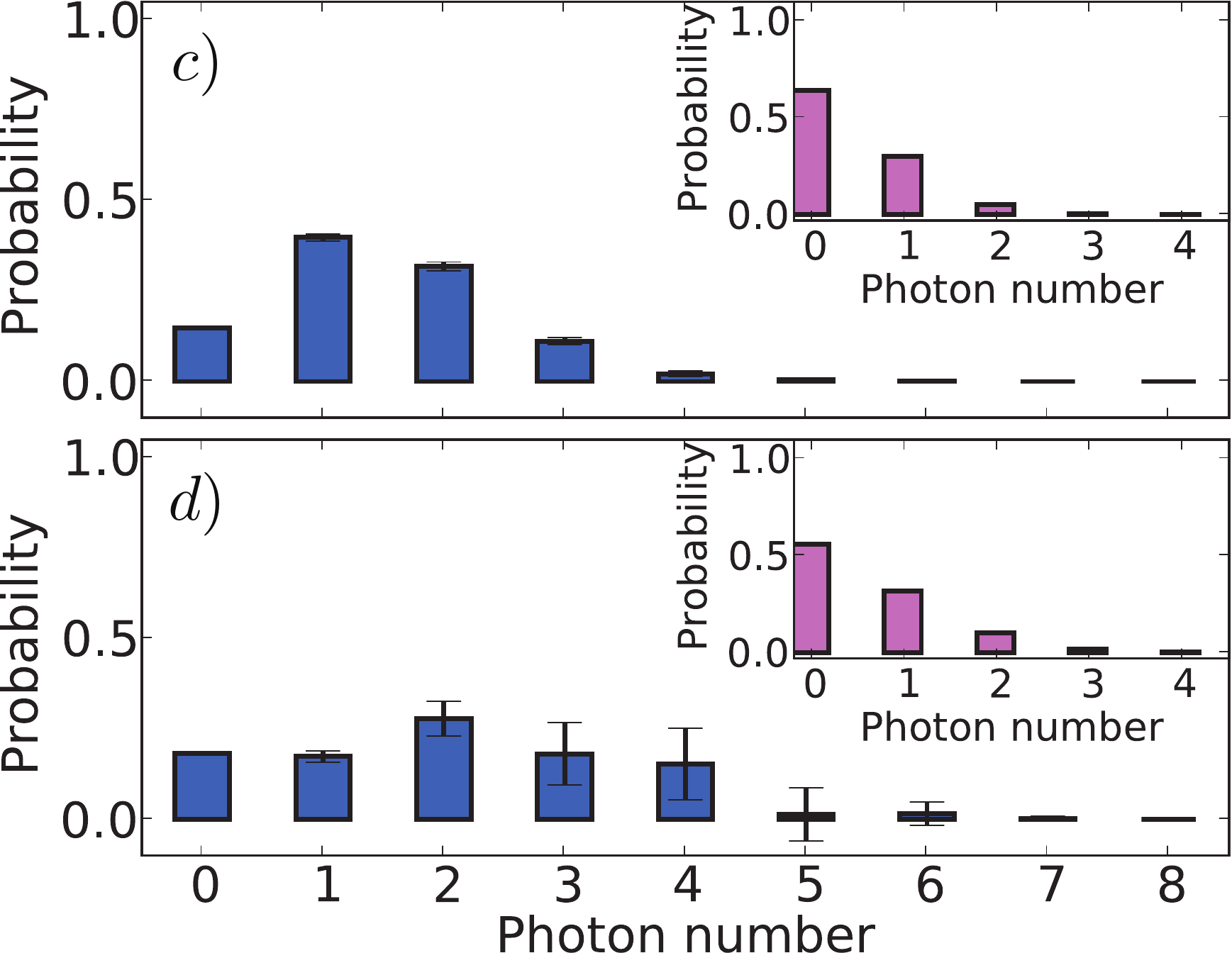}
	\caption{The loss-degraded (inset) and reconstructed photon number statistics of the single photon Fock state. The state is affected by 30\% loss before the displacement with 50\% overlap.. The following values of displacement were applied a) 0, b) 0.3, c) 0.7, d)1.0.}
		\label{30loss50overlap}
		\end{center}
\end{figure}

\begin{figure}[!ht]
	\begin{center}
		\includegraphics[width  = 0.45\textwidth]{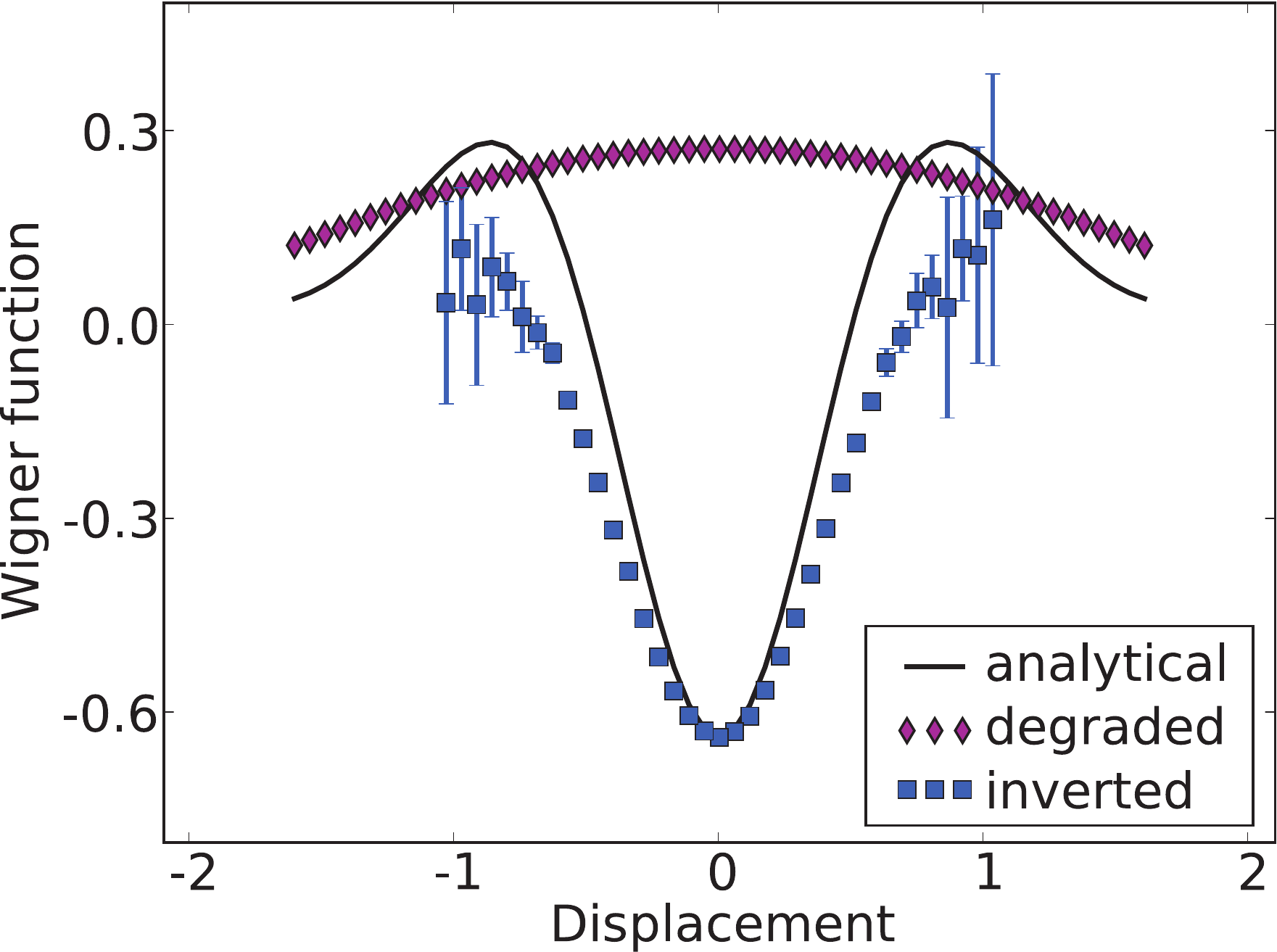}
		\caption{Loss-degraded (diamonds), reconstructed (squares), and analytical (solid line) Wigner functions for the single photon Fock state using the same parameters as in Fig.~\ref{30loss50overlap}.}
		\label{30loss50overlapW}
	\end{center}
\end{figure}

\paragraph{Case 4:}
Finally,  we  studied the reconstruction of the single photon Wigner function in the presence of imperfect mode overlap between the signal and reference fields. The statistics were generated according to Eq.~(\ref{P_convoluted}).  We now restricted the reconstruction range of the Wigner function to the region where probability components $\rho_{nn} > -0.003$. Other procedures, such as the error estimation, were kept the same. We utilised a beam splitter with 95\% transmittance for the implementation of the displacement and a an 8-bin TMD with ideal 50/50 couplers  and assumed that the single photon Fock state was affected by loss (efficiency equal to 30\%) before applying a non-ideal displacement ($\mathcal{M}= 0.5$).  In Fig.~\ref{30loss50overlap} we show the photon number statistics for both the degraded and reconstructed signals at different values of the displacement. 
 The quantum feature of  photon number oscillation is washed out by the dual action of losses and imperfect overlap. Nevertheless, the loss can still be handled by applying an appropriate inversion method. 
In terms of displacement, the reconstruction is reliable to the boundary value of $\alpha = 1.0$. 
As shown in Fig.~\ref{30loss50overlapW}, the Wigner function determined via the degraded signal does not display any quantum features. However, using the inversion of the statistics, the negative inner parts are recovered, although with a broadened width  as a consequence of the imperfect mode overlap.  
%

\section{\label{conclusions}Conclusions}

We studied the  reconstruction of  Wigner functions of low-photon-number Fock states in a TMD-based photon counting experiment.  
Several parameters restrict the reliable reconstruction range of the Wigner function in phase space:  the number of photons that a TMD can maximally resolve, the stability of the inversion  with respect to the level of loss and the amount of statistics collected, and the imperfect mode overlap.With current technology, the dominant limitation is the efficiency achievable by the experimental realisation.

Performing Monte Carlo simulations we reconstructed the Wigner functions for single-photon and two-photon Fock states at different levels of detection efficiency. We showed that, under perfect mode-matching conditions, the Wigner function of the single-photon Fock state can be reconstructed with high confidence. The characteristic oscillatory behaviour present in the photon number distributions of displaced Fock states are extremely sensitive to losses, yet this interesting quantum feature can be reliable restored by applying a loss-inversion. A clear indication of phase space interference is given by the suppression of the one-photon component of the displaced single-photon Fock state. 

Furthermore, we analysed the the situation of imperfect mode overlap.
We showed how to employ the  oscillatory behaviour as a tool to characterize the degree of mode overlap between the quantum signal and reference beam. Regarding the reconstruction of the Wigner function, even in a scenario of only 50\% mode overlap, the simulation shows the possibility of recovering the negative values of the single-photon Fock state, although with a broader width.

Our results motivate the experimental exploration of non-Gaussian states with photon number resolving detectors. This kind of experiments offers the opportunity of loss-tolerant {\it state} characterization via the inversion of losses,  along with the ability to probe the Wigner function point--by--point in phase space.
Furthermore, the probing method allows advanced quantum state characterization that is intrinsically not only sensitive to a single mode but takes into account all the characteristics of the original quantum state and the coherent reference field. 
In our paper, we provided the tools to reliably characterize the prepared state and highlighted the different impacts of inefficient detection, convolution of statistics and mode mismatch. 
In contrast to standard homodyne detection, our analysis enables us to recognise the type of experimental imperfections, and gives valuable information about the extent of degradation caused by each one. 
Hence, our work  constitutes a  basis for a broad understanding and efficient manipulation of experimental data. We expect that our investigation will open a new route of detecting and studying quantum states and offer a new experimental technique for hybrid quantum information applications. 

\section*{Acknowledgements}
 We would like to acknowledge fruitful discussions with T.~C.~ Ralph, W.~P.~Schleich and W.~Vogel. We thank P.~J.~Mosley  for his comments about the manuscript. This work was supported by the EU under QAP funded by the IST directorate as Contract No. 015848.

\appendix
\section{\label{App_A} Derivation of the Wigner function from loss-degraded statistics}

In this section we derive the analytical solution for reconstructing the Wigner function from the loss-degraded statistics. The elements of the loss matrix are determined by Eq.~(\ref{P_Nloss})
\begin{equation}
	L_{m,n} = \left \{ \begin{array}{l}\left( {\begin{array}{*{20}c} n \\ m \\ \end{array}} \right)\eta^{m}(1-\eta)^{n-m}, \quad \textrm{when}\quad m \le n \\
                         					0  \quad \textrm{otherwise},  \end{array} \right.  
	\label{loss}				
\end{equation} 
and the loss degraded statistics $p^{L}_{m}$ are connected to the  photon number statistics by
\begin{equation}
	p^{L}_{m} = \sum_{n = 0}^{N} \left( {\begin{array}{*{20}c} n \\ m \\ \end{array}} \right) \eta^{m}(1-\eta)^{n-m}\rho_{nn} , 
\end{equation}       
where  $N$ is the highest resolvable photon number of the detector,  $m$ is the number of detected photons while $n$ is the number of incident photons. Regardless of the efficiency ($ \eta \neq 0$), the direct inversion of the loss matrix is possible by
\begin{equation}
	L^{-1}_{m,n} = \left \{ \begin{array}{l}\left( {\begin{array}{*{20}c} n \\ m \\ \end{array}} \right)\frac{(-1+ \eta)^{n-m}}{\eta^{n}} , \quad \textrm{when}\quad m  \le n \\
                         					0 \quad \textrm{otherwise}, \end{array} \right .      
	\label{loss_inversion}				
\end{equation}
where $n$ now corresponds to the number of loss-degraded detection events and $m$ is the number of photons in the signal state.
Note that both matrices in Eqs.~(\ref{loss}) and (\ref{loss_inversion}) are upper diagonal matrices. 

The  inverted photon number statistics are given by
\begin{equation}
	\rho_{mm} = \sum_{n=0}^{N} \left( {\begin{array}{*{20}c} n \\ m \\ \end{array}} \right) \frac{(-1+\eta)^{n-m}}{\eta^{n}} p^{L}_{n}.
	\label{deg_pL}
\end{equation}
The reconstruction of the photon number $m$ is affected by all photon numbers  satisfying $n \ge m$. The  Wigner function  can be evaluated with the help of Eq.~(\ref{W_pn}) using the degraded statistics in Eq.~(\ref{deg_pL}) 
\begin{eqnarray}
	W & = &\frac{2}{\pi}\sum_{m= 0}^{N}\left(-1 \right)^{m}\rho_{mm} \nonumber \\
	   & = &\frac{2}{\pi} \sum_{n=0}^{N} \frac{1}{\eta^{n}} \left [ \sum_{m=0}^{n}\left( {\begin{array}{*{20}c} n \\ m \\ \end{array}} \right)(-1+\eta)^{n-m}\left( -1 \right)^{m} \right ] p^{L}_{m}, \qquad m \leq n \leq N \nonumber \\
  	  & = &\frac{2}{\pi} \sum_{n=0}^{N} \left( -\frac{2-\eta}{\eta} \right )^{n}  p^{L}_{n}.
  	  \label{Linv}
\end{eqnarray}
 The result of Eq.~(\ref{Linv}) agrees with the independent analysis presented in  Ref.~\cite{S.Wallentowitz1996} and corresponds to scaling the parity factor by the detection efficiency.

We can apply this method for example on the loss-degraded statistics of the single photon Fock state derived in Eq.~(\ref{P_displaced1}). The loss-inverted Wigner function can be written as 
\begin{eqnarray}
W &=& \frac{2}{\pi}\sum_{M}^{N\rightarrow \infty}\left ( -\frac{2-T}{T}\right ) ^{M}e^{-\left |\gamma\right |^{2}}\frac{|\gamma|^{2M}}{M!} \left [1-T+ \frac{T\left (M-|\gamma|^{2}\right)^{2}}{|\gamma|^{2}}\right ]\nonumber \\
& = & \frac{2}{\pi}e^{-2\frac{|\gamma|^{2}}{T}}\left [  4\frac{|\gamma|^2}{T} -1\right ],
\label{WFock1}
\end{eqnarray}
which  corresponds to the Wigner function of the single photon Fock state at the phase space point $\frac{\gamma}{\sqrt{T}}$ as expected. Experimentally,  the reconstruction of the Wigner function at any given point of phase space is not possible, since the maximal resolvable photon number upon detection is finite, i.e. for an $m$-stage TMD the summation upper bound $N = 2^{m}$ in Eq.~(\ref{Linv}). Nevertheless, concerning the characterization of low-photon number Fock states, the interesting regions of the Wigner function are around the origin of phase space and thus the probabilities for measuring high-photon number components are negligible. For these reasons, the reconstruction  range depends on the photon-number of the displaced state and on the resolution capability of the detector.

\vspace{3mm}

\end{document}